\documentclass[12pt,preprint]{aastex}

\newcounter{subfigure}

\shorttitle{On the polar caps of the 3 Musketeers}
\shortauthors{De Luca et al.}

\begin{document}

\title{On the polar caps of the Three 
Musketeers\footnote{Based on observations with XMM-Newton, an ESA science 
mission with instruments and contributions directly funded by ESA member
states and the USA (NASA).}}

\author{A. De Luca,  P. A. Caraveo, S. Mereghetti, M. Negroni\altaffilmark{1},
G.F. Bignami\altaffilmark{1,2}}
\affil{Istituto di Astrofisica Spaziale e Fisica Cosmica, \\
Sezione di Milano  ``G.Occhialini'' - INAF \\
v.Bassini 15, I-20133 Milano, Italy}
\email{deluca@mi.iasf.cnr.it}

\altaffiltext{1}{Universit\`a degli Studi di Pavia, Dipartimento di Fisica
Nucleare e Teorica, Via Bassi 6, 27100 Pavia, Italy}
\altaffiltext{2}{Centre d'etude spatiale des Rayonnements, CNRS-UPS, 
9, Avenue du Colonel Roche,  31028 Toulouse Cedex 4, France}

\begin{abstract}
XMM-Newton observations of PSR B0656+14, PSR B1055-52 
and Geminga have substantially increased 
the statistics available for these three 
isolated neutron stars, so apparently similar to 
deserve the nickname of ``Three Musketeers'' (Becker \& Tr\"umper, 1997). 
Here we shall take advantage of the EPIC statistics to perform 
phase 
resolved spectroscopy for all three objects.
The phase-averaged spectrum of the three musketeers is best described by
a three component model. This includes two blackbody components, 
a cooler one, possibly originating from the bulk of the star surface, and 
a hotter one, coming from a smaller portion of the star surface
(a ``hot spot''), plus a power law. 
The relative contributions of the three components
are seen to vary as a function of phase, as the stars' rotation bring
into view different emitting regions. The hot spots, which have very 
different apparent dimensions (in spite of the similarity of the 
three neutron stars polar cap radii)
are responsible for the bulk of the phase variation.
The amplitude of the observed  phase modulation is also markedly different  
for the 
three sources. Another striking aspect of our phase-resolved
phenomenology is the apparent lack
of any common phase alignement between the observed modulation
patterns for the two blackbody components. They are seen to 
vary {\em in phase} in the case of PSR B1055-52, but  {\em in anti-phase} 
in the case of PSR B0656+14.
These findings 
do not support
standard and simplistic models of neutron 
star magnetic
field configuration and surface temperature distribution.

\end{abstract}


\section{Introduction}
PSR B0656+14, PSR B1055-52  and Geminga are isolated neutron stars 
showing similar periods and 
period derivative. Hence their characteristic ages (few 100 thousand years), 
their inferred 
magnetic fields (few 10$^{12}$ G)
and their energetics ($\dot{E}_{rot} \sim 10^{34}$ 
erg s$^{-1}$) are 
very similar. Moreover, they are all nearby, making it easier to detect 
them at 
different wavelengths. Parallax distance measurements
are available for Geminga (157 pc, Caraveo et al., 1996) and PSR B0656+14
(288 pc, Brisken et al., 2003), while for PSR B1055-52 the current 
distance estimate, based
on the dispersion measure, is $\sim$750 pc (Kramer et al., 2003).
In X-rays they shine both owing to thermal and non-thermal mechanisms. 
Black-body emission, from the majority of their surface, is seen to be
complemented at higher energies by power law ``tails''.
They have faint optical counterparts, with 
visual magnitude m$_v $ $\sim$25.5 for
Geminga (Bignami et al., 1987), $\sim$25 for PSR B0656+14 (Caraveo et al., 
1994) and $\sim$24.9 for PSR B1055-52 (Mignani et al., 1997). Multicolor
photometry, available for Geminga and PSR B0656+14, shows that their NUV  
magnitudes are broadly compatible with the extrapolations of their X-ray 
blackbody emissions, while a power law-like excess is apparent in the 
optical/NIR (Mignani et al., 2004 and references therein).
While PSR B1055-52 and Geminga are bona fide gamma-ray pulsars (Fierro 
et al., 1993; Bertsch et al., 1992), 
for PSR B0656+14 a 
tentative detection  awaits confirmation (Ramanamurthy et al., 1996).

Here we shall concentrate on the analysis of the XMM-Newton
data collected both in imaging and timing mode.
Such exceptional harvest of X-ray photons over the ample energy interval 
covered by the EPIC cameras allows us to use a new tool for the study 
of the X-ray behaviour of  the ``Three musketeers'': 
phase resolved spectroscopy.

While we have already published
the phase-resolved analysis of the EPIC/pn data on Geminga (Caraveo et al. 2004),
here we shall include 
also 
this source,
in order to underline similarities 
and differences in the phase resolved X-ray behaviour of three otherwise very 
similar objects.

\subsection{Review of the X-ray observations of the Three Musketeers}
The EPIC data presented below come after two decades of X-ray observations of 
isolated neutron stars, starting with their discovery as soft, probably 
thermal sources by the Einstein Observatory.  
\subsubsection{Geminga}
1E 0630+178 was discovered by the Einstein satellite
while covering the gamma-ray error box 
of the unidentified source Geminga (Bignami et al., 1983). 
The discovery of a common 237 msec 
pulsation clicked the identification, the first ever, of an unidentified
gamma-ray source 
with its X-ray counterpart (Halpern \& Holt, 1992; Bertsch et al., 1992). 
The first good quality spectral data on Geminga 
were collected by ROSAT. Halpern and Ruderman (1993), using 7,911 photons in 
the 0.07-1.50 keV energy band, concluded that the spectrum was well described 
by two black-body curves at temperatures of 0.5$\times~10^6$ K and 
3$\times~10^6$ K. While the cooler component appeared to come from the full
surface of the neutron star, the hotter one could have been coming from a spot
covering just 3$\times10^{-5}$  of the neutron star surface, possibly due to a 
heated polar cap. A second, longer ROSAT observation, yielding 18,500 soft
photons, was analysed together with an observation by ASCA, providing the 
first 1,750 ``hard'' ($>$ 2 keV) counts. Such a combination seemed to favour a
different, composite interpretation, where, a power-law, non-thermal 
emission was added to the low temperature black-body radiation 
(Halpern and Wang, 1997).  The power-law photon index was still 
poorly constrained, 
ranging from 2.2 to 1.5.   Using a longer ASCA observation with 4,800 photons, 
Jackson et al. (2002) confirmed the composite nature of Geminga's spectrum and 
refined the power-law photon index value to  1.72$\pm$0.1.
They  tried also to perform some phase-resolved spectroscopy,
but could only use two phase intervals. They found the spectrum of the
interval 
including the peak to be slightly softer than the rest of the light curve, but 
concluded that ``the difference is only marginally significant''.

\subsubsection{PSR B0656+14}
The definitely brighter PSR 0656+14 was first detected by the Einstein 
satellite (Cordova et al., 1989). 
A ROSAT PSPC observation allowed Finley et al. (1992) to detect
the X-ray pulsation and to measure  a pulsed fraction of 14\% $\pm$ 2\%.
Further pontings were then carried out with   
the ROSAT PSPC detector in 1992
(Oegelman, 1995), for a total exposure time of about 
17 ksec, collecting $\sim$32,000 photons in the 0.1-2.4 keV band.
The overall ROSAT dataset was analyzed by Possenti et al. (1996), who
found the bulk of the emission to be of thermal origin, well described by 
a blackbody curve (T$\sim 9 \times 10 ^5$ K) originating from
a large part the star surface
(emitting radius $\sim$14 km,
assumig a distance of 760 pc,
corresponding to $\sim$5.3 km
at the parallactic distance of 288 pc).
A second spectral component was required to 
describe the higher energy part of the spectrum,
as well as to explain a significant change of the pulse profile with energy.
It was not possible however to 
discriminate between a second blackbody component (T$\sim 1.9 \times 10 ^6$ K)
originating from a region 
a few hundred times smaller
and a steep  ($\Gamma \sim 4.5$)
power law. 
Greiveldinger et al.(1996) coupled the ROSAT dataset to an ASCA 
observation which yielded $\sim$2,000 photons in the 0.5-10 keV band. 
Their best fit used a three component model: 
the sum of two blackbodies (best fit parameters
very similar to the results of Possenti et al. 1996) and of a power law with 
photon index $\Gamma$=1.5$\pm$1.1.
More recently, PSR B0656+14 was observed with Chandra, 
both with ACIS and with the LETG. The ACIS
data ($\sim$45,000 photons in the 0.2-6 keV range)
confirmed a three-component model to yield the best description of the
spectrum (Pavlov et al., 2002), consistent of the sum of two blackbodies 
(T$_1 \sim$0.85$\times 10^6$ K, 
R$_1 \sim$12 km; T$_2 \sim$1.65$\times 10^6$ 
K, R$_2 \sim$1 km, using a distance of 500 pc;  
assuming the distance of 288 pc,
R$_1 \sim$7 km and 
R$_2 \sim$0.6 km). 
The  high 
spectral resolution of the Gratings allowed Marshall \& Schultz (2002) to
exclude the presence of significant features superimposed on the thermal
continuum in the softer band (0.15-1 keV). The parameters
of their best fitting two-blackbody model are T$_1 \sim$0.8$\times 10^6$ K, 
R$_1 \sim$22.5 km; T$_2 \sim$1.5$\times 10^6$ 
K, R$_2 \sim$1.7 km, assuming a distance of 760 pc, 
corresponding to
R$_1 \sim$8.5 km and 
R$_2 \sim$650 m at the distance of 288 pc.

\subsubsection{PSR B1055-52}
After the Einstein observatory discovery 
of X-ray emission from this radio pulsar (Cheng \& Helfand, 1983),
PSR B1055-52 was observed with ROSAT, both with the HRI
(8.6 ksec yielding $\sim570$ source photons) and with the PSPC
(15.6 ksec, for a total of $\sim 5500$ source photons) in 1990-1992
(Oegelman \& Finley, 1993). 
The timing analysis unveiled the source pulsation, with a pulsed
fraction increasing from $\sim$11\% for energies $<$0.5 keV to
$\sim$63\% above 0.5 keV.
The spectrum was best described by a two component model. A 
blackbody with temperature of $\sim 8 \times 10 ^5$ K, coming from
a large portion of the surface, accounts for the bulk 
of the X-ray emission; a second component
was required, either a second, hotter 
(T$\sim 3.6 \times 10^6$ K)
blackbody coming from an area a few 10$^4$ times smaller, or a steep 
($\Gamma \sim 4$) power law.
An ASCA observation could add only $\sim$200 photons in the 0.5-10 keV range
(Greiveldinger et al., 1996).
A Chandra ACIS observation was taken in 2000 (42 ksec exposure); results were
summarized by Pavlov et al. (2002). A simultaneous fit to Chandra and ROSAT
data required a three component model, consisting of the sum of two
blackbodies 
(T$_1 \sim$8.3$\times10^5$ K, R$_1 \sim$12 km; kT
$_2 \sim1.6 \times 10^6$ K, R$_2 \sim$800 m assuming the distance to be 1 kpc; 
R$_1 \sim$9 km,
R$_2 \sim$600 m at the revised distance of
750 pc) 
and of a power law  
($\Gamma \sim$1.7). The above authors reported the presence of a significant
variation of the hot blackbody component, observed through phase-resolved
spectroscopy, but they did not present detailed results.  
A similar three component model has been used by Becker \& Aschenbach
  (2002) in their analysis of XMM-Newton data. Spectroscopy was performed
using MOS data only, which were fitted together with ROSAT data. Their 
best fit was obtained with T$_1 \sim$7.1$\times10^5$ K, R$_1 \sim$31 km, T
$_2 \sim1.4 \times 10^6$ K, R$_2 \sim$2.6 km assuming the distance to be 1 kpc
(at the revised distance of 750 pc the emitting radii are R$_1 \sim$15.5 km,
R$_2 \sim$1.3 km); the power law photon index was found to be 1.9$\pm$0.2
and the interstellar absorption N$_H\sim2.3\times10^{20}$ cm$^{-2}$.

\section{The XMM-Newton/EPIC data sets}

Geminga, PSR B0656+14 and PSR B1055-52 were observed by XMM-Newton
as Guaranteed Time targets. The complete journal of observations is reported
in Table~\ref{epicobs}. The data are now available in the XMM-Newton
Science Archive. 

While the MOS cameras (Turner et al., 2001) were operated in Full Frame
mode in order to image on the full field of view of
the telescopes (15$'$ radius), the pn detector (Str\"uder et al., 2001) 
was used to time-tag the photons, either in Small Window mode (6 ms time 
resolution, imaging on a 4$'\times4'$ field) 
or in Fast Timing Mode (0.03 ms resolution, no spatial information along
detector columns). 
Unfortunately, 
as pointed out by Becker \& Aschenbach (2002),
the use of the potentially promising Fast Timing mode produces,  
as a byproduct, a background significantly higher than in Small Window mode,
owing to the peculiar readout mode and to the collapse of data along the CCD
columns. 
Such an abnormally high background 
lowers the
signal-to-noise above a few keV.


\subsection{Data reduction and background screening}
\label{datared}
All the data reduction was performed using the most recent release of 
XMM-Newton Science Analysis Software (SASv6.0.0). The raw Observation Data 
Files (ODFs) were processed using standard pipeline tasks ({\em epproc} for pn,
{\em emproc} for MOS data). We selected events with PATTERN 0-4 and PATTERN 
0-12 for the pn and the MOS, respectively.

Particular care was devoted to reduce the
instrumental background in the linearized event lists.
First, an accurate screening for soft proton flare events was done, 
following the prescription of De Luca \& Molendi (2004).
We computed Good Time Intervals (GTIs) setting a 
threshold of $3\sigma$ from the quiescent rate on the 
0.5-12 keV count rate for both the pn and the MOS detectors.
A more stringent threshold  was adopted in the case of pn
Fast Timing mode observations, since the collapse of data along the CCD readout
direction increases (by a factor $\sim30$) 
the background count rate in the source extraction region. 

The pn Fast Timing mode data required a further cleaning.
Such operating mode is affected by a peculiar flaring background component
(``Soft Flares'', see Burwitz et al., 2004) ultimately due to the interaction
of charged particles (possibly heavy ions) with the CCD.
Such interactions produce short
(0.1-0.5 s time scale) and very intense bursts of events, with typical
energies of $\sim0.22$ keV for singles (mono-pixel events) and of $\sim 0.45$ 
keV for doubles (bi-pixel events). 
To remove such background noise, which would 
hamper the study of the low-energy emission of our targets, ad-hoc GTI 
files for singles and doubles were created, extracting 1-sec binned light 
curves in the 0.2-0.3 keV and 0.4-0.5 keV energy ranges, 
respectively, and then using the same algorithm adopted for the soft 
proton case.

The overall GTI filter for each pn Fast Timing observation was 
then obtained by intersecting the soft proton GTI, the soft flares GTI
for singles and the soft flares GTI for doubles.     
The net exposure times (after dead-time correction) for the cleaned event
lists are reported in Table~\ref{epicobs}. In the case of PSR B1055-52
the observation was splitted between two XMM revolutions (186 and 187). 
Event lists obtained with the same instrument in the same mode in different
orbits were then merged using the SAS task {\em merge}.

\subsection{Source and background events selection}
\label{srcbkg}
To extract the source photons from the dataset taken
in imaging modes (Full Frame for MOS and Small Window for pn) 
we selected a circle of  45$''$ radius, containing $\sim90$\% of the counts;
in Fast Timing mode we used a 17 pixel wide strip (4.1$''$ pixel size),
containing $\sim85$\% of the source counts. Background photons were extracted
from suitable regions on the same CCD chip containing the source: for
Full Frame MOS images we need an annulus of 2$'$ inner radius 
and 4$'$ outer radius; for the pn detector operated in small window 
we selected two rectangular regions oriented along the CCD readout direction
and covering $\sim$2 square arcmin; for pn used in Fast Timing mode 
we selected two stripes 
(17 and 12 pixels wide) away from the source region.

The 0.25-8.0 keV count rate observed in MOS1 camera for PSR B0656+14 is
of $\sim0.9$ counts s$^{-1}$. A modest pile-up is expected, owing to the
slow CCD readout in the Full Frame mode (2.6 sec frame time). An analysis 
of the event PATTERN distribution, performed with the SAS task 
{\em epatplot}, clearly showed such effect as an excess of bi-pixel events
above 0.7 keV. To solve the problem, we simply excluded the PSF core (inner
5$''$, containing about 35\% of the source photons), where the pile up 
could be 
significant. No hints for pile-up in the resulting photon list was apparent
in the corresponding event PATTERN distribution.

In the case of PSR B1055-52, EPIC MOS images show a faint source 
at an angular distance of 32$''$ in the NE direction
(see Figure 17 of Becker \& Aschenbach, 2002).
Such source has a hard spectrum (possibly a background AGN);
its flux, negligible wrt. the pulsar below 2 keV, becomes comparable 
to the pulsar's one above 3 keV. In order
to avoid possible contaminations in the study of the pulsar high energy
emission we decided to exclude such source from the pulsar extraction
region, using a 10$''$ arcsec radius circle (containing $\sim 60$\% of the
counts at 5 keV) as a geometrical mask. Such a selection was not possible in
pn Timing mode, since along the RAWX direction (perpendicular to the 
readout direction), where spatial information is mantained, the angular
separation of the two sources is $<$4 pixels. Therefore, the flux measured
by the pn above $\sim$3 keV is possibly contaminated by some contribution from
such background source and its absolute value should be taken with caution.
 
Background-subtracted count rates for the three neutron stars are reported
in Table~\ref{epicobs}, together with the number of collected photons. 

\section{EPIC Data analysis}

With a number of time-tagged photons (see Table~\ref{epicobs}) more than 
doubling all previous statistics, 
EPIC offers now the first 
chance of meaningful phase-resolved spectroscopy for the three musketeers.

Following the procedure outlined by Caraveo et al. (2004): 
\begin{itemize}
\item first, we address the phase averaged spectral shape of the three objects 
to obtain useful pieces of information on the components needed to fit their 
overall spectra;
\item next, we perform the timing analysis to search for the best period 
and to perform the phase alignement;
\item then, we divide the folded light-curves in 10 phase intervals and extract
 the spectra corresponding to each phase interval;
\item finally, each spectrum is fitted using the same components found to best 
describe the integrated spectrum and the corresponding phase-resolved spectral
parameters are computed.
\end{itemize}

In the following section we will discuss in detail each 
step for the cases of PSR B0656+14 and PSR B1055-52. The case of Geminga
was described by Caraveo et al.(2004); we will report here the results to
ease a synoptic view of the phenomenology of the three musketeers.  

\subsection{Phase-integrated spectral analysis}
\label{phaseint_spectroscopy}
Spectra for the source and background events, extracted using the regions
described in Sect.~\ref{srcbkg}, were rebinned in order to have at least
40 counts/channel. We added a 5\% systematic error to each spectral bin,
to account for calibration uncertainties among different instruments/modes.
 {\em Ad hoc} redistribution matrices and effective area
files were generated using the {\em rmfgen} and {\em arfgen} tasks of the SAS.
The spectral analysis was performed with XSPEC v11.2. 
We selected 0.25 keV as lower bound for the spectral study, since below
such an energy the calibration is still uncertain; the upper bound
was selected on the basis of the observed signal-to-noise, 
which depends both on the sources' spectra and on the operating mode used.
As discussed in Sect.~\ref{datared},
the spectra collected by the pn in Fast Timing mode suffer from 
a rather
high background, which hampers the study of our targets at high energy, where
such sources are rather faint. 
For PSR B0656+14 spectra obtained in Fast Timing mode turned out to be useful up to 2
keV, while for PSR B1055-52 we could include data up to 6 keV.
Spectra from MOS1/2 Full Frame mode, as well as from pn sw mode, were used up to 8 keV.
MOS and pn spectra were
fitted simultaneously, leaving a cross-normalization factor as the only free
parameter.

\begin{figure}
\includegraphics[angle=-90,width=15cm]{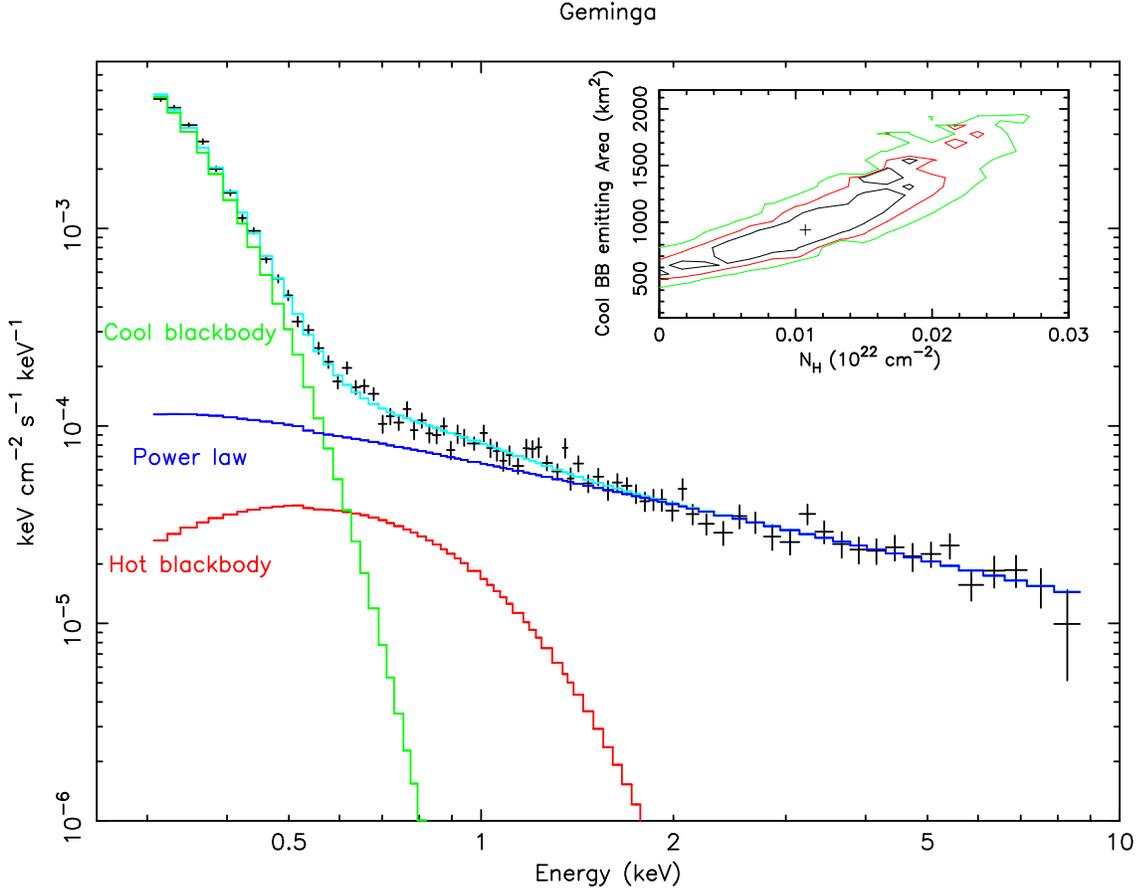}
\caption{\label{spectrum_geminga} 
Unfolded phase-integrated
spectrum of Geminga. 
Only data from pn camera are  plotted.
This figure is adapted from Caraveo et al.(2004); a different color code
is used here (see also Fig.~\ref{spectrum_0656} and Fig.~\ref{spectrum_1055}).
The best fitting spectral model
is represented by the light blue line. As discussed in the text, this is
obtained by the sum of a cool blackbody component (green), a hot blackbody
component (red) and a power law (blue). Detailed values of the best
fitting  parameters are reported in Table~\ref{spectralres}.
The inset shows confidence contours for the interstellar column density
  N$_H$ vs. the emitting surface for the cool blackbody. 68\%, 90\% and 99\%
confidence levels for two parameters of interest are plotted. Caraveo et
al.(2004) fixed the N$_H$ value to 1.07$\times10^{20}$ cm$^{-2}$ (resulting
from ROSAT data) to obtain their set of best fitting parameters 
(see Table~\ref{spectralres}).}

\end{figure}

\begin{figure}
\includegraphics[angle=-90,width=15cm]{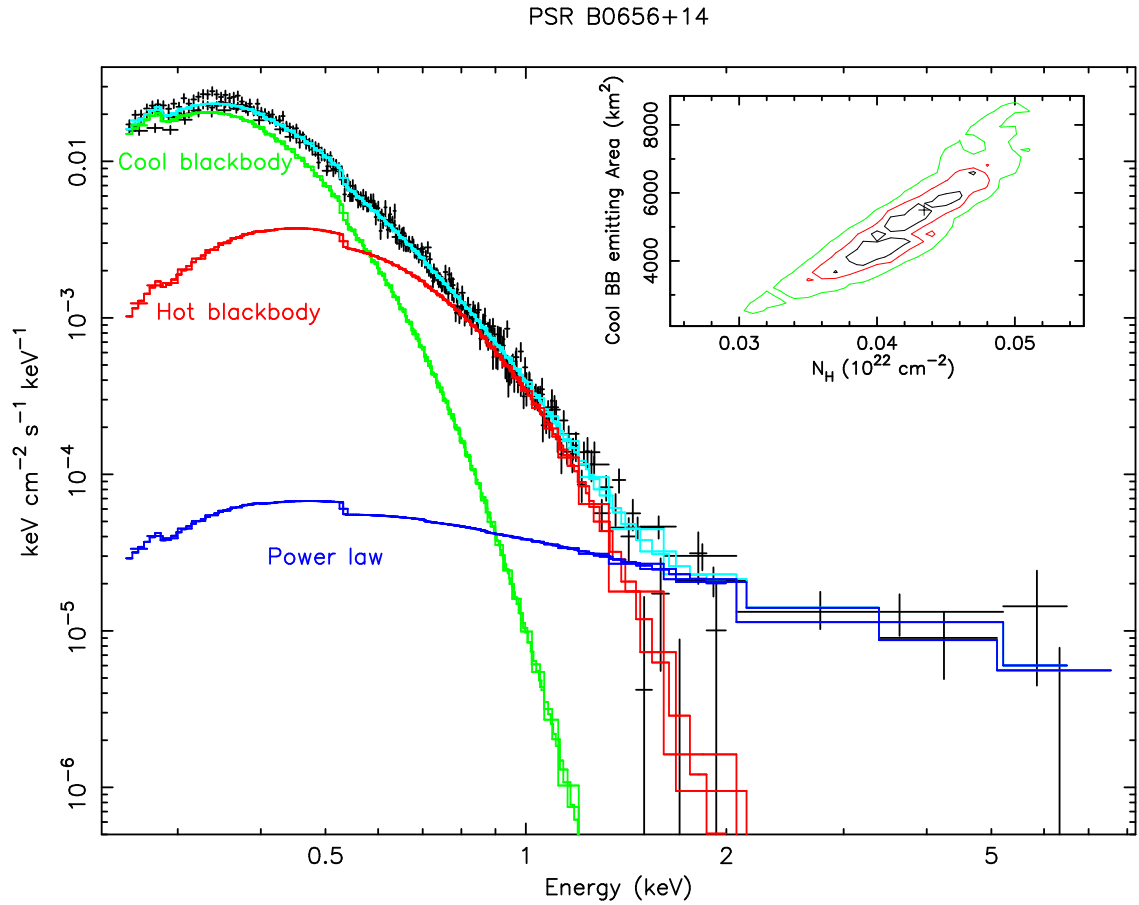}
\caption{\label{spectrum_0656} Same as Fig.~\ref{spectrum_geminga}
for the case of PSR B0656+14.
Data from pn (both Small Window and Fast Timing
mode) and MOS1 are plotted (black points). 
Detailed values of the best
fitting parameters are reported in Table~\ref{spectralres}.
}
\end{figure}

\begin{figure}
\includegraphics[angle=-90,width=15cm]{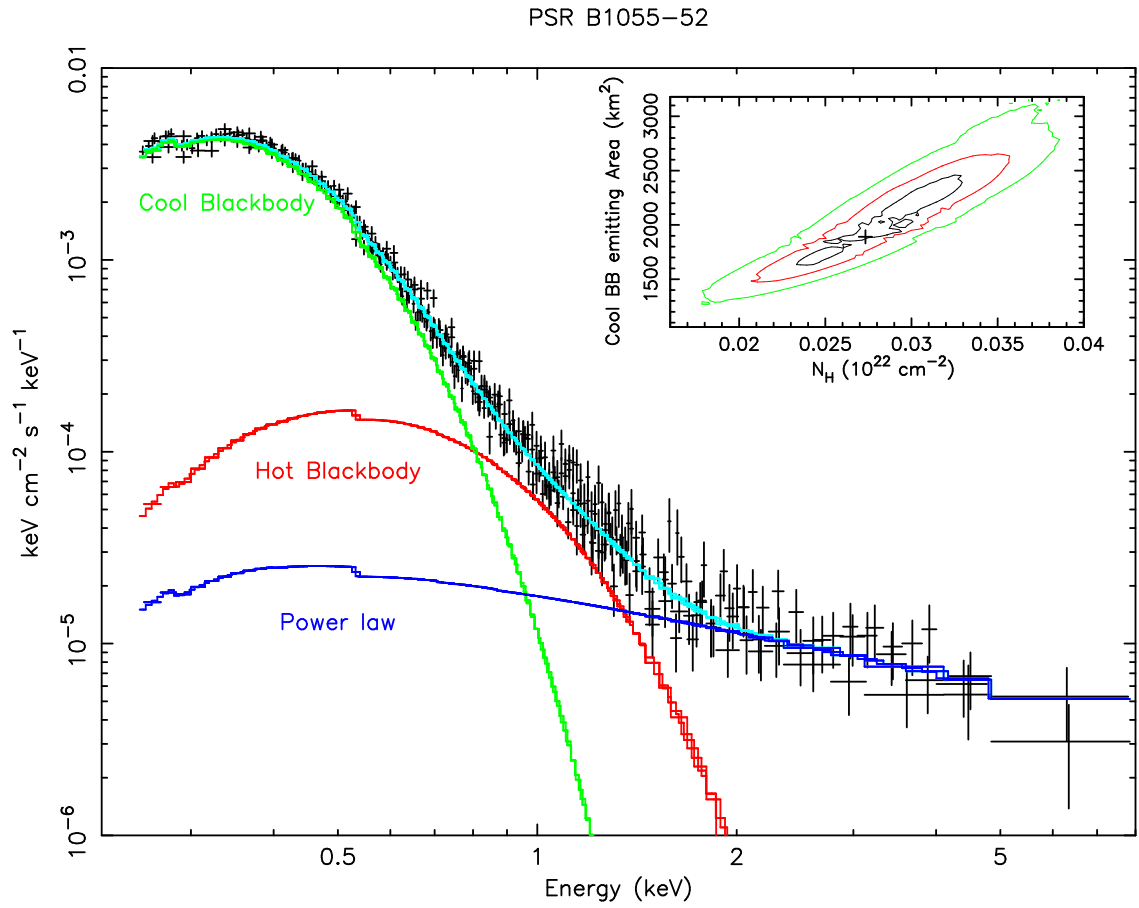}
\caption{\label{spectrum_1055} 
Same as Fig.~\ref{spectrum_geminga}  
for the case of PSR B1055-52. Data
are from pn, MOS1 and MOS2. See Table~\ref{spectralres} for
details on the best fitting spectral parameters.}
\end{figure}

No significant spectral features are detected superimposed on the continuum of
the three musketeers, 
neither in emission nor in absorption.
As in the case of Geminga (Caraveo et al., 2004) (Fig.\ref{spectrum_geminga}),
the best fitting model is found
combining two blackbody curves and a power law.
This confirms the findings
of Pavlov et al. (2002), based on Chandra observations of
PSR B0656+14 and PSR B1055-52, and Becker \& Aschenbach (2002), 
who analyzed  XMM-Newton (MOS only) data
for PSR B1055-52.
Using the best fitting parameters reported in Table~\ref{spectralres} 
we obtain $\chi^{2}_{\nu}$=1.1 (368 d.o.f.) for PSR B0656+14
(Fig.\ref{spectrum_0656}) and 
$\chi^{2}_{\nu}$=1.0 (327 d.o.f.) for PSR B1055-52 (Fig.\ref{spectrum_1055}).
The use of a second power law instead of the hotter blackbody
yields photon indices of $\sim6.5$  for PSR B0656+14 
and $\sim5.6$ for 
PSR B1055-52, while lowering the fit quality ($\chi^{2}_{\nu}>$1.5 and
$\chi^{2}_{\nu}>$1.1 for the two cases, respectively).
We note that a single-temperature, magnetized
 neutron star atmosphere model (Zavlin et al. 1996) 
cannot adequately reproduce the profile of the low-energy part of 
the spectrum of PSR B0656+14 and PSR B1055-52, similarly to the 
case of a single 
blackbody curve.  Moreover, such model requires a very
large emitting surface (with emitting radii of $\sim$100 km - this is true
also for Geminga). Therefore such model 
cannot provide a satisfactory description of the spectra 
of the three musketeers.

\begin{figure}
\includegraphics[angle=-90,width=15cm]{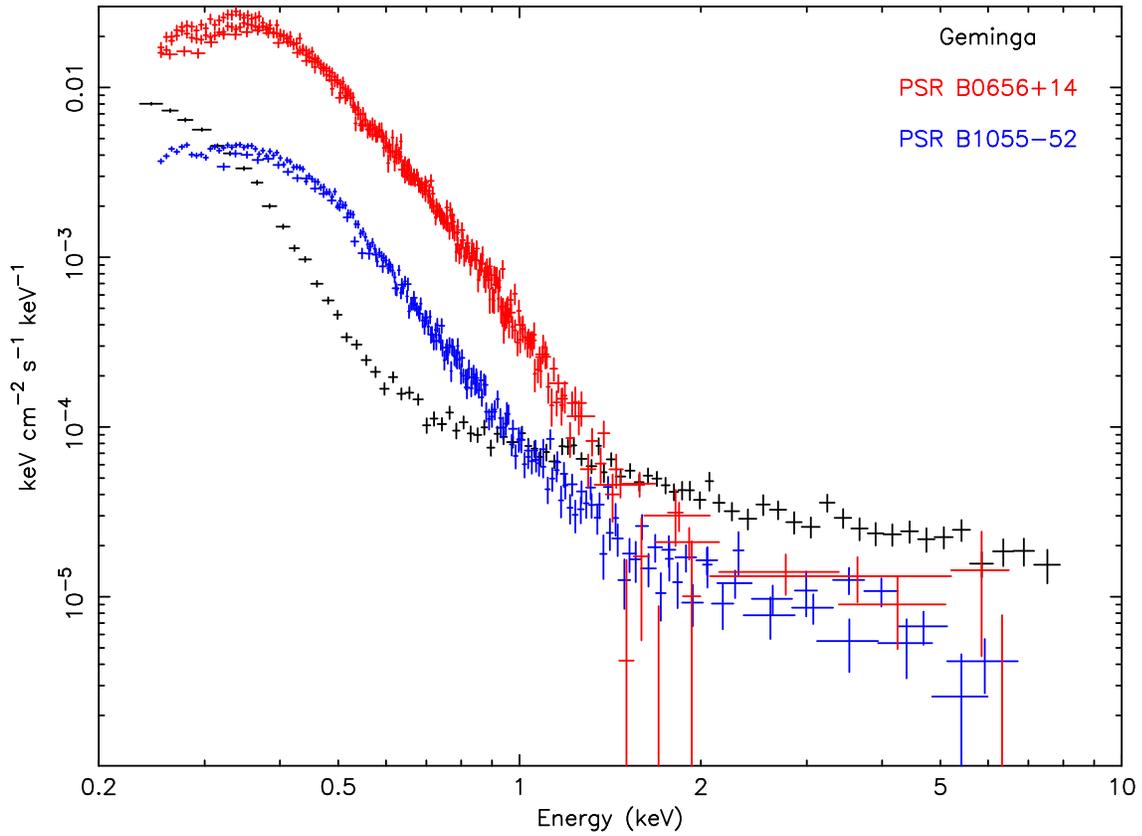}
\caption{\label{spectra} Unfolded spectra of PSR B0656+14 (red, 
data from pn and MOS1), PSR B1055-52
(green, data from pn, MOS1 and MOS2) and Geminga (black, data from pn). 
See text for details. }
\end{figure}

A synoptic plot of the
spectra of the three musketeers is shown in Fig.~\ref{spectra}.
It is easy to note the definitely higher
signal to noise in the high energy portion ($>$2 keV) of the spectrum 
of Geminga, in spite of a flux only slightly higher in such energy range.
This is mainly due to 
the different operating modes used for the pn detector in the
observations of the three targets, especially to
the much lower background in the pn small window mode 
wrt. Fast timing mode. The power law component in the spectra of PSR B0656+14
and PSR B1055-52 is constrained by the MOS spectra and 
(in the case of PSR B0656+14) by the short pn small
window observation 
(see the caption to Fig.~\ref{spectra} for 
further details).
We summarize the results of the EPIC spectral fits for the three musketeers
in Table~\ref{spectralres}, 
where errors are computed at 90\% confidence
level for a single interesting parameter.
The lower temperature blackbody (hereafter ``cool blackbody'') is associated
to an emitting region with an area compatible with the full surface
of the neutron star. The higher temperature component (``hot blackbody'') 
is seen to originate from a much smaller area. 
To visualize the correlation between different spectral parameters we computed
confidence contours (at 68\%, 90\% and 99\% levels for two parameters
of interest)
for the interstellar column density N$_H$ vs.
the cool blackbody emitting area;
this is shown in the insets in Fig.~\ref{spectrum_geminga}, 
Fig.~\ref{spectrum_0656} and Fig.~\ref{spectrum_1055}.

As shown in Fig.~\ref{spectra},
the overall shape of the spectrum of the three musketeers is very similar.
However remarkable differences in the luminosity of different
spectral components are immediately evident (see Table~\ref{spectralres}). 
In particular, we note that the luminosity of the hot blackbody in the case
of Geminga is $\sim2$ orders of magnitude lower than in the other two cases. 
At variance with PSR B0656+14 and PSR B1055-52, for Geminga 
the contribution of the
power law dominates over the hot blackbody in the energy range where the
hot blackbody has its maximum.


The parameters best fitting the EPIC spectrum of PSR B1055-52 (see Table~\ref{spectralres})
are fully consistent with the results reported by Becker \& Aschenbach (2002) and 
by Pavlov et al.(2002), who included ROSAT data in their spectral fits to EPIC/MOS and
Chandra/ACIS data, respectively. 
The same is true in the case of Geminga: 
although we use a 3 component model, the results for the Cool blackbody component 
(temperature and emitting radius) are found
to be very similar to the ROSAT ones (see e.g. Halpern \& Ruderman, 1993).
The case of PSR B0656+14 is slightly different: 
the interstellar column density ($4.3\pm0.2\times10^{20}$ cm$^{-2}$) 
resulting from the symultaneous fit to the pn/Fast Timing, pn/Small Window and MOS1/Full Frame data is
larger than the values ($\sim 2\times10^{20}$ cm$^{-2}$) 
obtained by 
Pavlov et al.(2002) and Marshall \& Schulz (2002) (based on Chandra ACIS/LETG data)
and by Possenti et al.(1996) (based on ROSAT data).
Thus, our analysis of the EPIC data results in a brighter
underlying continuum at low energies,
leading to a lower temperature coupled to a larger emitting surface for PSR B0656+14.
Addressing such a discrepancy, possibly due to subtle cross-calibration
issues, is beyond the scope of our fitting exercise, which was
performed to provide the starting point for the phase-resolved
spectroscopy of the EPIC data.


\subsection{Timing analysis}
As  a first step, we have studied the high time resolution pn data in order
to derive the period of our targets. 
Source  photons were selected from the regions described above, in the energy
range 0.15-8 keV for pn Small Window mode and 0.15-2 keV for pn Fast Timing
mode\footnote{Time-tagging of  photons is reliable over the whole detector
sensitivity range; therefore we include in the timing analysis also low
energy  (E$<$0.25 keV) photons, not used for spectroscopy (see 
Sect.\ref{phaseint_spectroscopy})}. 
Photon time  of arrivals were
converted to the Solar System barycenter using the SAS task {\em barycen} 
and folded  using 10 bins in a range of trial periods around the 
expected values.
A very significant  detection of the pulsation was obtained in each case.
To determine with  higher accuracy the period value and to evaluate
the corresponding error, we followed the prescription of Leahy (1987).

The results are reported in Table~\ref{timingres}, where they are 
compared with
the extrapolations of the radio ephemeris, selected 
to be the nearest to
the XMM-Newton observation. The period for 
the radio-quiet Geminga is compared to the $\gamma$-ray ephemeris
of Jackson et al. (2002), based on the analysis of the complete EGRET dataset.
A perfect consistency between 
the expected and measured values is apparent. 

\begin{figure}
\includegraphics[angle=-90,width=15cm]{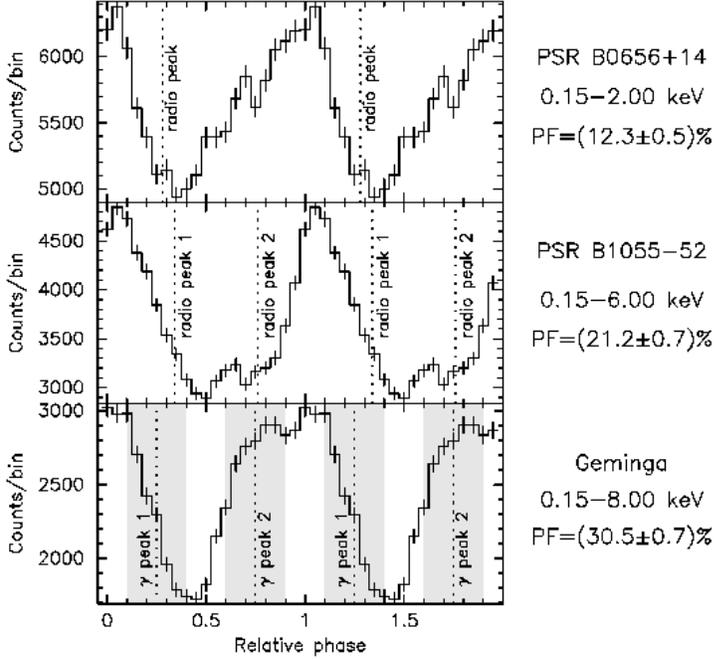}
\caption{\label{lctot} Light curves of the three musketeers. Data from
pn (energy ranges: 0.15-2 keV for PSR B0656+14; {\bf 0.15-6} keV for PSR B1055-52;
0.15-8 keV for Geminga)
have been folded using the radio timing solutions reported in 
Table~\ref{timingres} for PSR B0656+14 and PSR B1055-52 and the 
EGRET $\gamma$-ray ephemeris (also reported in Table~\ref{timingres}) for
Geminga. The phase has been set in order to put the X-ray maximum at phase 0.
The phases of the radio peaks have been marked with vertical dashed lines;
their uncertainty is estimated to be $\sim$0.01 (1/5 of phase bin) for the case of PSR
  B0656+14 and $\sim$0.003 ($\sim$1/20 of phase bin) for PSR B1055-52. See
  text for further details.
For PSR B1055-52 ``radio peak 1'' refers to the highest peak in the
radio profile, see also Fig.~\ref{lc_1055}. 
We plotted also the phases 
of the $\gamma$-ray peaks for Geminga; however, as discussed by Caraveo et 
al.(2004), the propagation of errors in the extrapolation of the EGRET 
ephemeris makes their position uncertain by $\pm$0.15 ( 
such a phase interval corresponds to the gray-shaded regions; ``$\gamma$ peak
1'' refers to the highest peak in the $\gamma$-ray profile, see also 
Fig.~\ref{lc_geminga})}
\end{figure}

In order to align in phase the X-ray light curves with the radio (or gamma) ones, 
we have folded XMM-Newton data using the radio timing
solutions for PSR B0656+14 and PSR B1055-52 and the gamma one for Geminga.
The absolute accuracy of the XMM-Newton clock 
is $\sim 500~\mu s$, as stated by the
calibration team (Kirsch, 2004) and by recent investigations  (e.g. Becker et al., 2004).
The extrapolation of the radio ephemeris to the time of the XMM observations
yields a phase uncertainty of $\sim$0.01 for PSR B0656+14 (corresponding to
1/5 of phase bin in Fig.~\ref{lctot})
and of $\sim$0.003 (1/20 of phase bin in Fig.~\ref{lctot}) for PSR
B1055-52, for which the clock accuracy is the
dominant factor, since the radio data are almost simultaneous to the X-ray ones.

A larger uncertainty, $\sim0.15$ in phase, is present in the case of 
Geminga, as discussed by Caraveo et al. (2004), owing to the much longer time 
separation between the EGRET and XMM-Newton observations.

The folded pulse profiles (background-subtracted)
of our targets using the overall EPIC
energy range are shown 
in Fig.~\ref{lctot}.
To ease their comparison, all the lightcurves have been plotted 
setting phase 0 to
the X-ray pulse maximum as seen in the overall energy band. 
The phases of the radio peaks (PSR B0656+14 and PSR B1055-52) and
$\gamma$-ray peaks (Geminga and PSR B1055-52) are also shown. 
Pulsed fractions have been computed as the ratio between the counts above
the minimum and the total number of counts, 
deriving the corresponding uncertainties from 
the propagation of the statistical errors in the folded light curves.

\begin{figure}
\includegraphics[angle=-90,width=15cm]{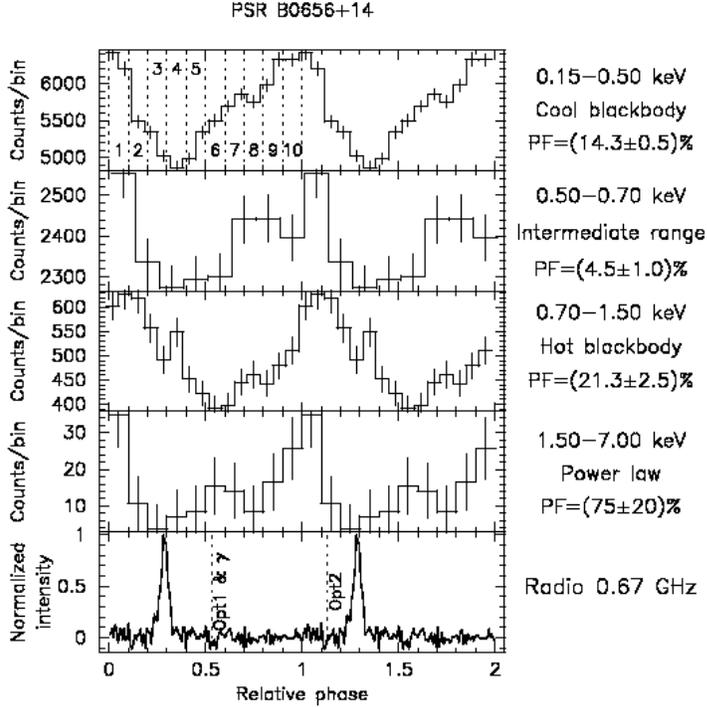}
\caption{\label{lc_0656} 
Lightcurves of PSR B0656+14 in different energy ranges. Data 
(obtained with pn Fast Timing mode observations, 
with the exception of the hardest
band, based on pn Small Window mode data) 
have been
folded using the radio ephemeris quoted in Table~\ref{timingres}. 
The alignement in phase is the same chosen for Fig.~\ref{lctot}.
Although always single peaked, the pulse profile changes significantly
going from
the softest energy range (dominated 
by the cool blackbody) to the 0.7-1.5 keV range (dominated by the hot 
blackbody). Note the minimum in the pulsed fraction in the intermediate 
0.5-0.7 keV range. Above 1.5 keV the lack of statistic hampers a detailed 
study of the pulse profile. The radio light curve is shown in the lower
panel. The uncertainty on the phase alignment of the X-ray light curve
with the radio one is of $\sim$0.01.
The single radio pulse is seen to trail the X-ray maximum 
by $\sim0.2$ in phase. The phases of the two optical peaks (Kern et al., 2003) as well as of
the $\gamma$-ray peak claimed by Ramanamurthy et al.(1996) are also
shown. See text for further details.}
\end{figure}

\begin{figure}
\includegraphics[angle=-90,width=15cm]{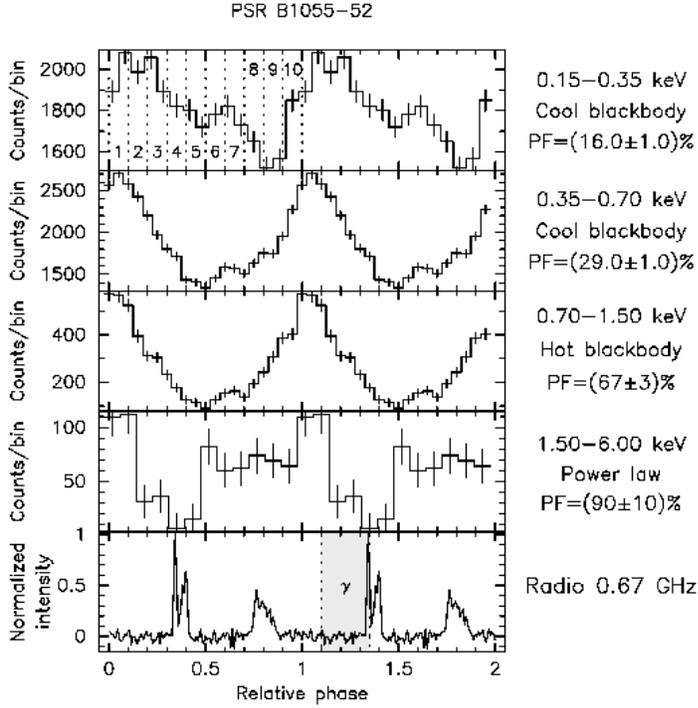}
\caption{\label{lc_1055}
Light curves of PSR B1055-52. Note
the broader profile of the X-ray pulse below 0.35 keV and the
small ($\sim0.1$) phase shift wrt. higher energies. 
Note the value of the pulsed fraction, which
grows with energy and is remarkably high ($\sim70$\%) in the 0.7-1.5 keV range,
where the neutron star emission is dominated by the hot blackbody
component. For a more detailed study of the energy resolved pulsed fractions see
Becker \& Aschenbach (2002).
Above 1.5 keV the low signal-to-noise hampers a detailed
study of the pulse profile. We plotted in the lower panel the radio
light curve. The uncertainty on the phase alignment is of $\sim0.003$.
The two radio peaks are observed to bracket the single
X-ray peak. The $\gamma$-ray pulse (Thompson et al., 1999) occurs in the
gray-shaded phase interval
between the two vertical dashed lines.}
\end{figure}

\begin{figure}
\includegraphics[angle=-90,width=15cm]{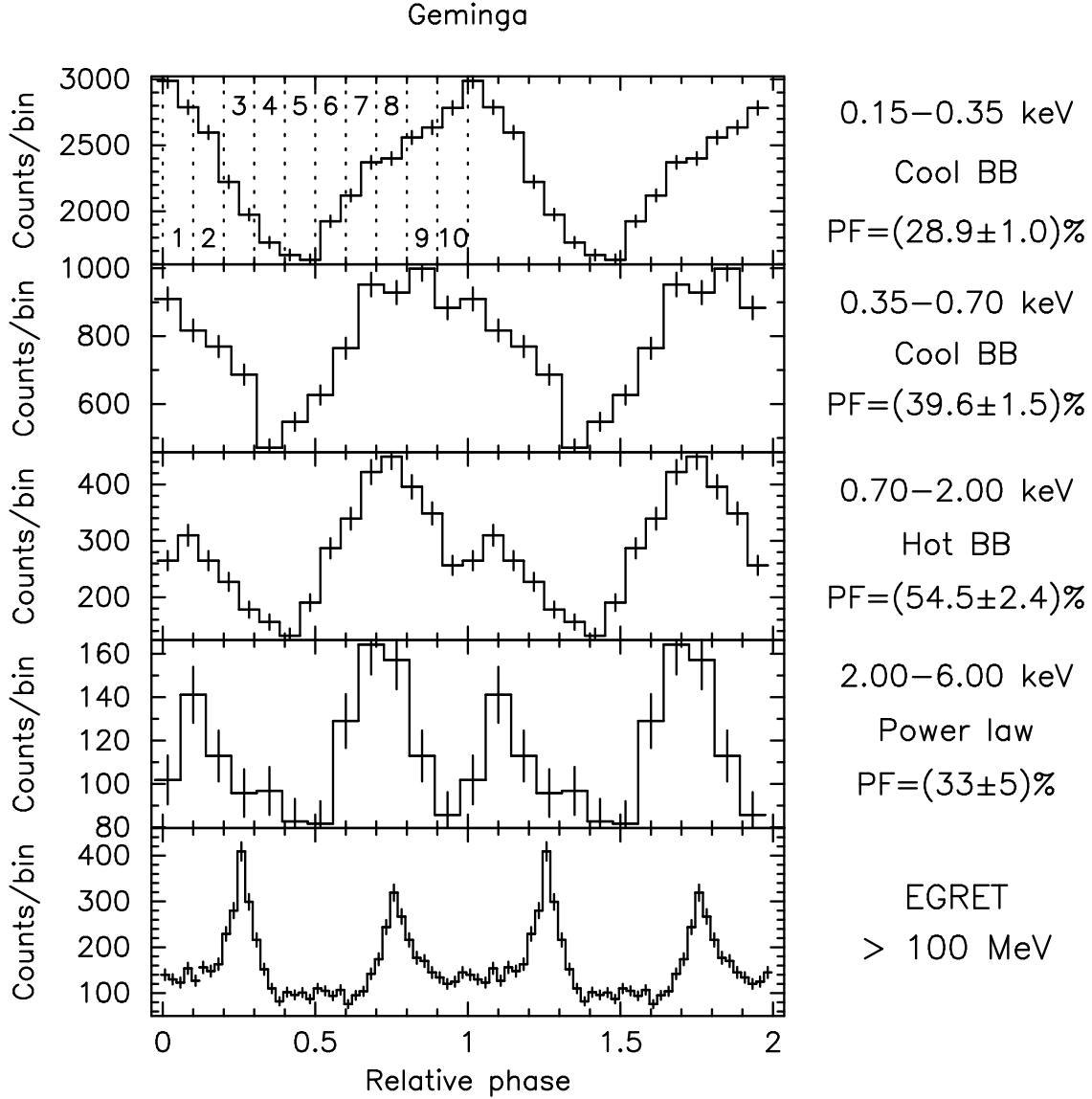}
\caption{\label{lc_geminga} 
Light curves
of Geminga. As discussed by Caraveo et al.(2004), 
the pulse shape changes as a function of energy. The single, broad peak
observed at low energy (where emission from the cool blackbody dominates)
changes to two peaks at higher energies (where the power law component
dominates). The pulsed fraction is maximum in the 0.7-2 keV range, where
the hot blackbody component is more important. The lower panel show the 
EGRET $\gamma$-ray lightcurve. The extrapolation of the $\gamma$-ray ephemeris
makes the absolute phase alignement uncertain by $\pm$0.15,
owing to the long time span between the EGRET and the EPIC observations.
Note that the numbering of the phase intervals defined in the upper
panel is different from that used by Caraveo et al. (2004).
Their phase 1 is  here phase 3 and similarly for all.}
\end{figure}

We have then studied the variations of the pulse profiles with energy.
The light curves relative to four energy intervals are shown in 
Fig.~\ref{lc_0656} (PSR B0656+14), Fig~\ref{lc_1055} (PSR B1055-52) 
and Fig.~\ref{lc_geminga} (Geminga). Note that the energy
ranges used are different for different targets, 
since for each source the energy intervals have been 
chosen in order to visualize the light curve relative 
to the cool blackbody, transition region, hot blackbody
and power law. 
The lower panels of figures ~\ref{lc_0656} and ~\ref{lc_1055}
show the radio profile of PSR B0656+14 and PSR B1055-52 at 0.67
GHz\footnote{Retrieved from the European Pulsar Network 
database, http://www.mpifr.bonn.mpg.de/div/pulsar/data/browser.html}.
For the case of Geminga (Fig.~\ref{lc_geminga}), the $\gamma$-ray light 
curve\footnote{this was obtained by selecting 9 EGRET Viewing periods 
in which the target was imaged close to the center of the 
field of view and having a good photon statistics; 
events with energy E$>$100 MeV were extracted from the source position;
time of arrivals were converted to the solar system barycenter
and folded with the ephemeris published by Jackson et al. (2002)} is shown.
Ten phase intervals,
selected {\em a priori}, to be used for phase-resolved spectroscopy
(see Sect.~\ref{phaseres_spectroscopy})
are marked in the upper panel.
Our three neutron stars exhibit vastly different phenomenologies.
The main results may be summarized as follows (see also the captions
to Fig.~\ref{lc_0656}, Fig~\ref{lc_1055} and Fig.~\ref{lc_geminga}):

\begin{itemize}
\item For PSR B0656+14 a single pulse per period may be seen in each
energy band. The pulse shape below 1.5 keV is broadly sinusoidal;
a significant change in pulse profile occurs between
the energy range dominated by the cool blackbody and the range 
dominated by the hot blackbody. As a consequence, in the intermediate 
range the pulsed fraction reaches its minumum value. At higher energy,
the pulse profile seems sharper, although the low statistics does not allow
to study its shape (nor the pulsed fraction) with great accuracy. Note that the light curve above
1.5 keV was obtained by folding data from the short pn Small Window mode
observation; the other cases represent data from the longer Fast Timing 
observation. 
Considering the overall light curve (Fig.~\ref{lctot}),
the single radio peak lags the X-ray maximum  
by $\sim0.25\pm0.05$ in phase
(the uncertainty 
is dominated by 
the X-ray light curve bin width for 
the determination of the position of the X-ray peak),
occurring very close to the X-ray minimum.
In the lower panel of Fig.~\ref{lc_0656} we mark the phases of the 
double-peaked pulse profile observed in the V band by Kern et al.(2003).
The position of the single $\gamma$-ray peak 
claimed by Ramanamurthy et al.(1996) is also indicated. 

\item Also PSR B1055-52 shows a single pulse per period at all energies,
with a markedly different shape in the softest band (below 0.35 keV) 
wrt. intermediate energy ranges (0.35-1.5 keV), while above 1.5 keV the low 
signal-to-noise hampers the study 
of the pulse profile. 
The pulsed fraction, which is seen to grow with energy, is remarkably
high ($\sim70$\%) in the band where the hot blackbody dominates the neutron
star emission.
The two peaks of the radio profile are seen to contour the X-ray peak at all energies;
in the overall light curve the highest radio peak lags 
the X-ray one by 
$\sim0.2\pm0.05$ (the uncertainty is dominated by the 
X-ray light curve binning).
The phase interval where the $\gamma$-ray pulse occurs
lies between the X-ray maximum and the highest radio peak.

\item Geminga shows a different
behaviour of the pulse shape as a function of energy. 
The single, broad peak seen
at low energy, where the cool blackbody dominates, changes to a double
peak towards higher energies, where the bulk of the emission is non-thermal
(power law component). The pulsed fraction reaches a maximum of $>$50\% in the
energy range where the hot blackbody is more important.
The significant uncertainty ($\sim0.15$) affecting the extrapolation of the
EGRET ephemeris (see above) prevents any firm
conclusion on the alignement between the X-ray 
and the $\gamma$-ray light curves.

\end{itemize}

\subsection{Phase-resolved spectroscopy}
\label{phaseres_spectroscopy}
The  results of the spectral and timing analysis were then used to study
the variation of the emission spectrum with pulse phase.
For  both PSR B0656+14 and PSR B1055-52
we selected photons whose times of arrival fall into the 10 different  
phase  intervals marked on Fig.~\ref{lc_0656} and Fig.~\ref{lc_1055}.
The time resolution of the pn Fast Timing mode (0.03 ms) and Small Window
mode (5.6 ms)  are adequate to perform such photon phase selection.
We then
extracted 10  phase-resolved spectra. For the case of PSR B0656+14 this was
independently done for both Fast Timing data and Small Window data.
Each spectrum  was then rebinned in order to have at least 40 counts per bin.
Since now we are interested to study {\em relative} variations as a function
of the pulsar  phase, we didn't add systematic errors to the pn 
phase-resolved spectra.

As already suggested by  the analysis of the energy-resolved pulse profiles,
the spectra of both sources do vary 
as a function of the  pulsars' rotational phase. 
This is easily seen  in the first 
column of Fig.~\ref{phaseres_0656_a} and ~\ref{phaseres_0656_b} 
(PSR  B0656+14),  
Fig.~\ref{phaseres_1055_a}  and ~\ref{phaseres_1055_b}
(PSR B1055-52), where the observed phase-resolved
spectra are compared to the best fit model describing the phase-averaged
ones and the deviations are computed in units of statistical standard errors
(an animated version of Figures 9a/b, 10a/b and 11a/b is available at
http://www.mi.iasf.cnr.it/$\sim$deluca/3musk/).
The variation is dramatic 
in the case of PSR B1055-52, owing to the high phase modulation of its
emission (up to 70\% in the 0.7-1.5 keV range, see Fig.~\ref{lc_1055}) 
while it is less spectacular (but still seen
with high significance) for PSR B0656+14, due to its lower modulation
(pulsed fraction $\leq$20\% below 2 keV, see Fig.~\ref{lc_0656}).
Note that in the case of PSR B0656+14 phase-resolved spectra 
have been rebinned in order to ease 
the visibility of the modulation on the graphs' scale.
PSR B0656+14 and PSR B1055-52
seem therefore to share the same characteristics seen in
Geminga, but their behaviours follow different trends
(the case of Geminga is reported in Fig.~\ref{phaseres_geminga_a}
and ~\ref{phaseres_geminga_b}).

Turning now to the spectral fit of the phase-resolved spectra,
no meaningful results could be obtained by allowing both the temperature 
and the normalization of the blackbodies to vary:
a definite trend showing the maximum temperature
in correspondence of minimum emitting area (and vice-versa) was systematically obtained
in all cases. Such a modulation pattern is obviously driven by the strong correlation between the 
two spectral parameters, and cannot be trusted as reliable.

Thus, we have
used the phase-integrated best fit model as a template to describe the 
phase-resolved data. The N$_{H}$ value,
the temperatures of both blackbodies and the power law photon index were fixed 
to the values best fitting the phase-integrated spectrum. The phase 
modulation was then reproduced by allowing the normalization parameters
of each spectral component to vary independently. 
For the case of PSR B0656+14, 
spectra from Small Window data and from Fast Timing data corresponding to the 
same phase interval were fitted simultaneously.

As in the case of Geminga (Caraveo et al., 2004), such a simple approach 
(using only three free parameters) yielded a satisfactory description
of the phase-resolved spectra also for PSR 
B0656+14 and PSR B1055-52. Reduced $\chi^{2}$ of $\sim0.8-1.4$ were obtained
in different phase intervals.  

The results of phase-resolved spectral fits are shown in right column of 
Fig.~\ref{phaseres_0656_a} and ~\ref{phaseres_0656_b},
Fig.~\ref{phaseres_1055_a} and ~\ref{phaseres_1055_b} 
(the  case of Geminga is  represented in
Fig.~\ref{phaseres_geminga_a}
and  ~\ref{phaseres_geminga_b}): for each phase interval, the three 
best  fitting spectral
components are plotted, superimposed
to  the unfolded data points. 
Knowing the objects' distance values, we can then  compute
the  emitting surface corresponding to each blackbody curve.
To visualize the evolution of the spectral parameters as
a function of the pulsars' phase, we have plotted the
blackbody radii, as well as the power law intensities,
as a function of the objects' rotational phase.
This  is shown in 
Fig.~\ref{parms_0656}, Fig.~\ref{parms_1055} and Fig.~\ref{parms_geminga}.




The  behaviour of the three neutron stars as a 
function of their rotational phase is vastly different.
The  most important results may be summarized as follows:
\begin{itemize}
\item The cool  blackbody component shows a similar
phase evolution for the three targets.
The values of  the blackbody radii follow a roughly sinusoidal 
profile with a modulation $\leq$10\% (wrt. the average value) for PSR B0656+14 
and  for PSR B1055-52,
of  $\sim$15\% for Geminga. 
\item  The hot blackbody components have a similar phase evolution with
a sinusoidal profile. 
Hovever,  we see striking differences both in the amplitude of their
phase modulation and on their overall luminosity.
While for  PSR B0656+14 the modulation
in the emitting radius wrt. the average value is $<$10\%,
similar to  the value found for the cool blackbody component,
in the case of PSR B1055-52 we see a 100\% modulation, 
since the hot  blackbody component is not seen in 4 out of 10 
phase intervals.
A  similar, 100\% modulation is observed also in the case of Geminga, 
although in this case the hot blackbody component is seen to disappear
in just  one phase interval, and the profile of its phase
evolution is markedly broader.
As far as  the radii values are concerned, we note that they span a factor
of 30, ranging from the 60 m of Geminga to the $\sim$2 km of PSR B0656+14.
\item The study  of the power law component is 
partially hampered, in the cases of
PSR B0656+14  and PSR B1055-52, by the low signal-to-noise at energies above
1.5 keV, owing to the high background affecting the Fast Timing observations.
It is therefore  difficult to assess the shape of the pulse profile 
(single-peaked for PSR B0656+14?; double-peaked for PSR B1055-52?) and the
actual modulation in such cases.  Conversely, for Geminga, 
thanks to the different pn mode, the power law 
component show a clear  double-peaked profile, with a significant
unpulsed flux.
\item  The relative phase evolution of the three spectral components is 
vastly different for the three neutron stars. This is particularly evident
when looking  at
the two blackbody components. In the case of PSR B0656+14 some sort of
anti-correlation  is observed, with the hot blackbody peaking in correspondence
of the cool blackbody minimum. This is at odds with the behaviour observed
for PSR B1055-52, where the two thermal  components appear definitely
correlated.
Geminga  shows an intermediate phenomenology: the cool and the hot
blackbodies have a $\sim0.25$ difference in their phase pattern.
\end{itemize}

Of course, one could 
fit the phase-resolved spectra
fixing the emitting radii to the average values and 
leaving the temperatures as free parameters.
The goodness of the fits 
obtained following such an approach is comparable
to that obtained previously fixing the temperatures and the photon index.
Not surprisingly, the evolution
of the parameters, as a function of the phase, is very similar to the plots shown
in Fig.~\ref{parms_0656}, Fig.~\ref{parms_1055} and Fig.~\ref{parms_geminga}.
However, such an approach renders less immediate the interpretation
of the hot blackbody phase evolution especially in the case of PSR
B1055-52, where the hot spot is not present for about half of the period. 
Maximum emitting radius translates into maximum temperature;
absence of hot spot translates into T(hot blackbody)=T(cool blackbody),
implying important changes of temperature for the same polar cap,
indeed a small portion of the neutron star surface.

\section{Discussion}
The phase-resolved spectroscopy of PSR B0656+14,
PSR B1055-52 and  Geminga allows for a new 
view of the
phenomenology of these middle-aged neutron stars.
While  the phase-averaged spectra of our targets look very similar, their
phase-resolved behaviour is quite different.
This  is particularly evident from 
Fig.~\ref{parms_0656}, Fig.~\ref{parms_1055}, and Fig.~\ref{parms_geminga},
where the  evolution of the cool and hot blackbody emitting radii, as well as of the 
power law flux, is shown as a function of the pulsars' phase.
Owing to the  not optimal quality of data collected for PSR B0656+14 and 
PSR B1055-52 above 2 keV, here we shall concentrate on the analysis of the 
two thermal  components, presumably coming from the star surface,
albeit from different regions at different temperatures.
\subsection{The cooler component}
\label{cbb}
First,  let us examine the cooler component. It is seen to come from a region
encompassing a good fraction, if not the totality, of the neutron star surface
as a  result of the star cooling.
We note that the best fitting emitting surface in the case of PSR
  B0656+14 
is very large 
(observed emitting radius of $\sim$ 21 km) if compared with expectations 
for a standard neutron star.
Distance uncertainties cannot ease the problem, owing to the very accurate
radio VLBI parallax measurement (Brisken et al., 2003).
In any case, a value of 15 km, corresponding to a somewhat stiff 
equation of state (see Lattimer \& Prakash, 2001 and references therein), is marginally compatible with data (being allowed
within the 99\% confidence contour plot for N$_H$ and the emitting surface shown 
in Fig.~\ref{spectrum_0656}).
The pulsed fraction observed in the 
energy range where the cooler component dominates is of 14\% for
PSR B0656+14 (Fig.~\ref{lc_0656}), 
of 16\% for PSR B1055-52 (Fig.~\ref{lc_1055}), while for Geminga it is higher
than 30\% (Fig.~\ref{lc_geminga}). 

We ascribe such a modulation to a variation 
of the emitting areas,
going from a sizeable fraction to the totality of the
neutron star surface as a function of the rotational phase. 
A viable mechanism to provide some sort of phase-dependent ``obscuration''
of the bulk of the neutron star surface could be the magnetospheric
 ``blanket'', originally described by Halpern \& Ruderman (1993) in their
study of the soft thermal emission from Geminga
(see also Halpern \& Wang, 1997): cyclotron resonance 
scattering by plasma in the magnetosphere at a few stellar radii could
screen the thermally emitting surface during specific phase intervals,
depending on the magnetic field configuration and viewing geometry.

Alternatively, the flux variations
could also be due to large-scale surface temperature modulations,
expected as a consequence of anisotropic heat transfer from the neutron
star interior (Greenstein \& Hartke, 1983). 
As discussed in Sect.~\ref{phaseres_spectroscopy}, such an interpretation
is consistent with the observed spectral phase variation as well.
With the current data 
we are not able to disentangle
temperature and emitting surface variation without a complete physical model.
We note that the observed pulsed fraction values are much greater 
than expected on the basis of
simple thermal models. As shown by Page et al. (1995), a few \% modulation
is expected in large-scale surface thermal emission, owing to the effects of
gravitational bending. Pulsed fractions as high as the observed values 
could be explained
assuming a peculiar beaming of the thermal surface emission. 
Owing to anisotropic
radiative transfer in a magnetized plasma,
an anisotropic
angular emission pattern wrt the surface's normal is indeed expected, 
depending on temperature and magnetic field intensity and
configuration\footnote{However, as noted in Sect.~\ref{phaseint_spectroscopy}, 
current atmospheric models do not yield a satisfactory description of the 
spectra of the three musketeers} (e.g. Zavlin \& Pavlov 2002 and
references therein). 
\subsection{The hot spot(s)}
Turning now to the hotter component, we note that for all objects this is the 
most dramatically variable spectral component. It is natural
to interpret such marked variations as an effect of the star rotation, which 
brings into view and hides one or more hot spots on the star surface.
It is commonly accepted that neutron stars should have polar caps hotter than
the rest of the surface. This could be due to
different processes such as the bombardment of charged 
particles
accelerated in the magnetosphere and falling back to the polar caps along
magnetic field lines (return currents, see e.g. Ruderman \& Sutherland 1975;
Arons \& Scharlemann 1979),
 or anisotropic heat transfer 
from the neutron star core, which depends strongly on the magnetic field 
direction and is maximum along the magnetic field lines (see e.g. 
Greenstein and Hartke 1983). 
An association of the observed rotating hot spot(s) with 
the neutron stars' polar caps
seems therefore rather obvious. 
\subsubsection{Luminosity and size}
The observed hot spot bolometric luminosities vary by more than 2
orders of magnitude, from the rather similar 
values found for PSR B0656+14 and PSR B1055-52
(5.7$\times 10^{31}$ erg s$^{-1}$ and  1.6$\times 10^{31}$ erg s$^{-1}$,
respectively), 
to the much dimmer Geminga (1.6$\times 10^{29}$ erg s$^{-1}$).
The former two cases are in broad agreement with the theoretical 
expectations of Harding \& Muslimov (2002) for polar cap heating due to
downstreaming of e$^{+/-}$ generated by curvature radiation photons. 
In the case of the Geminga pulsar, close to the
death line for creation of e$^{+/-}$ couples via curvature radiation, 
the lower polar cap
luminosity is consistent with expectations  
for polar cap heating due to bombardment by
particles created by inverse Compton scattered photons only 
(Harding \& Muslimov, 2002; see Caraveo et
al., 2004).

Straight estimates of neutron star polar cap size, based on a simple 
``centered''  dipole magnetic field geometry 
(polar cap radius R$_{PC}=R\sqrt{\frac{R \Omega}{c}}$), where R is the 
neutron star radius, $\Omega$ is the angular frequency and c is the speed of
light),  predict very similar radii for the three neutron stars, 
characterized by similar periods 
(233 m for PSR B0656+14, 326 m for PSR B1055-52 and 297 m for Geminga,
assuming a standard neutron star radius of 10 km). 
The observed emitting radii
are instead markedly different, with values ranging 
from $\sim 50$ m for Geminga
to $\sim$2 km for PSR B0656+14. 
Psaltis et al.(2000) showed that the observed polar cap radius may be
different from the actual one by a large amount, due to geometrical effects.
Indeed, Caraveo et al.(2004) proposed a highly inclined viewing
angle to reduce the surface of the emitting region in the case
of Geminga. 
PSR B0656+14 and PSR B1055-52 face a completely different situation,
since their
polar caps are significantly larger than expected.
Standard estimates, based on pure geometrical
considerations, are clearly unsatisfactory.
Even using the unrealistically large neutron star radius found for
PSR B0656+14, the inferred polar cap dimension is $<$600 m, far less
than the hot spot radius.
It is also possible, as suggested by Ruderman (2003), that
thermal photons be significantly reprocessed higher up in the magnetosphere, 
interacting with charged particles. 
In such a picture, the same mechanism possibly providing the bulk of the
cool blackbody modulation (via the phase-dependent screening quoted
in Sect.~\ref{cbb}) would obviously
bias all emitting radii (and temperatures) measurements based on blackbody fits.
\subsubsection{Modulation with phase}
The observed phase modulation of the hot blackbody component is largely
different for the three neutron stars both in amplitude and in pattern.
In the case of PSR B1055-52 a 100\% modulation is seen,
the hot spot disappears for 4/10 of the pulsar period.
Such phenomenology strongly
argues for an oblique rotator seen at high inclination. 
This would favour the historical Rankin (1993) interpretation
of the radio polarization pattern of this pulsar 
(orthogonal rotator) versus the older
one (almost aligned rotator)
of Lyne \& Manchester (1988); however, the presence of only one
visible ``pole'' would represent a challenge for a classical orthogonal
rotator seen at high inclination.
Conversely, the much
lower modulation observed in the case of  PSR B0656+14 seems consistent
with an almost aligned rotator, with the polar cap always in sight.
This picture is in good agreement with the radio pulse polarization results
of Everett \& Weisberg (2001), as well as with the above mentioned studies
of Rankin (1993) and Lyne \& Manchester (1988).
The remarkably high pulsed fraction of PSR B1055-52 
($\sim$70\% in the 0.7-1.5 keV range) requires a significant
beaming of the hot thermal component emission. Indeed, Psaltis 
et al.(2000) estimated
that under standard assumptions about the star mass and equation of state,
pulsed fraction higher than 35\% cannot be produced even in the 
most ``optimistic'' case of an orthogonal rotator having very small polar caps
with a high temperature contrast wrt. the rest of the
star surface. The same consideration applies to the case of Geminga
(pulsed fraction of $\sim55$\% in the 0.7-2 keV range, but with a significant
contribution of non-thermal emission).
In all cases a single peak per period is observed. This may
suggest that we are seeing a single hot region on the surface 
of the star. The visibility of a single pole along the star rotation
is consistent with the geometrical interpretation of
the phase-resolved behaviour of PSR B0656+14, while for PSR B1055-5
an important role of beaming should be invoked to explain why
the opposite pole emission is not seen, in spite of the
effects of gravitational light bending (Zavlin et al., 1995).
An alternative possibility could be a magnetic field configuration different
from a standard centered dipole. 
In the case of a magnetic field with significant multipole components,
two (or more) hot polar caps could be very close on the star surface 
and their emission could be blended in a single peak.
\subsection{Phase alignement between the thermal components}
Owing to
the sensitivity of heat transfer to the magnetic field, a temperature
anisotropy should exist
on the neutron star surface, with temperature increasing towards 
the magnetic poles (e.g. Greenstein \& Hartke, 1983).
Assuming the angular distribution of the radiation emitted by a surface element to be
peaked along the normal to the surface and decreasing towards larger angles
(see Harding \& Muslimov 1998 and references therein),
we would expect the overall emission of the neutron star
to be modulated as the poles come in and out of our line of sight.
In particular, maximum flux should be observed when the line of sight 
is best aligned with the hotter regions.
If the hot spots are indeed at the magnetic poles, a definite correlation between the 
hotter and colder components should be clearly visible in the phase
evolution of the emitting regions. 
While this is indeed the case for PSR B1055-52, the contrary is 
true for PSR B0656+14, which shows a clear anti-correlation.
Geminga shows an intermediate phenomenology, 
with a phase difference of
$\sim$0.25 between the hot and cool components 
(Fig.~\ref{parms_geminga}). 
Thus, one (or more) of the assumptions within the above
simple scenario (temperature distribution
resulting from a centered dipole magnetic field; pencil beaming) are not correct.
Different hypotheses should be considered:
the surface temperature distribution is possibly more complicated,
e.g. as a consequence of a multipolar magnetic field; the hot 
spots may not be located close to the center of the hotter surface
region; the emission beaming function may have strong peaks at angles
away from the normal to the surface element.
We note that magnetospheric reprocessing of thermal photons (see previous
sections) could possibly
ease the problem. Such a mechanism would introduce phase delays between thermal
spectral components which would not be directly related to the properties
of the surface, but driven by the magnetospheric structure.


The observed phase-resolved behaviour of the three musketeers
does not seem to argue in favour
of the simple, ``canonical'' picture of 
neutron stars as inclined, rotating, centered dipoles. 

\section{Conclusions}
In general, the new X-ray phenomenology revealed by the EPIC/XMM-Newton
combination presents new aspects of these three isolated, local neutron stars.
Rotating, polar hot spots are clearly detected for the first time.
It is tempting to link their origin to energetic particle bombardment,
following the Ruderman mechanism, especially because certainly two
(and possibly all) of the three musketeers are strong $\gamma$-ray
sources. In the case of Geminga, moreover, Caraveo et al. (2003)
have found a truly ``smoking gun'' evidence for this: the object's bow shock
is filled with electrons of just the right energy ($\sim 10^{14}$ eV) and
luminosity (a few $10^{28}$ erg s$^{-1}$) to be the particles escaping
from a polar cap.

However, the measurement of polar cap sizes, rendered accurate by the
excellent distance measurements, does not match simple theoretical 
expectations. 3-D geometry, 
beaming physics as well as the role of magnetospheric scattering will have to
be invoked for detailed models. The same holds true for the apparent
sizes of the cool blackbody components. At least in one case they 
exceed the  expected size 
of a standard neutron star. Simple atmospheric models worsen the problem.
However, before considering the implications of this finding on neutron
stars equations of state, we must understand if the parameters of the
neutron stars surfaces can be reliably derived from their thermal
radiation or if some other mechanism is biasing our results.

Finally, the apparent puzzle posed by the difference in the cool and hot 
components phase 
alignements in the three objects is difficult to reconcile with existing
pictures. This is reminescent, however, of the situation at $\gamma$-ray
energies, with different light curves. It will be very interesting to
look for contemporary X-ray and $\gamma$-ray data when AGILE and GLAST
will be in orbit with XMM-Newton.

\acknowledgments
We thank R. Manchester for his help with the Parkes radio ephemeris and 
A. Possenti for his useful suggestions about radio 
timing data. We thank A. Pellizzoni and M. Conti for their help with the 
EGRET data on Geminga.
The XMM-Newton data analysis is supported by the Italian Space Agency (ASI).
ADL acknowledges an ASI fellowship.

\clearpage

\begin{table}
\begin{center}
\caption{\label{epicobs} Journal of XMM-Newton observations.} 
\begin{small}
\begin{tabular}{cccccccc} 
\hline \hline
Pulsar/Date/Obs.Time & Camera(mode)\tablenotemark{a} & Good Time & Energy & Photons(\%bkg) & Count rate \\ \hline
PSR B0656+14 & pn(SW) & 5970 & 0.15-8.0 & 44600(1.7\%)  & 7.34$\pm$0.05 \\ \cline{2-6}
  2001-10-23 & pn(Ti)  & 16850 & 0.15-2.0 & 120000(6.3\%) &6.67$\pm$0.04  \\ \cline{2-6}
   41.0 ksec     & MOS1(FF) & 37800 & 0.15-8.0 & 28100(2.1\%) & 0.728$\pm$0.007 \\ \hline \hline
PSR B1055-52 & pn(Ti) & 61900 & 0.15-6.0 & 84450(14.4\%) & 1.167$\pm$0.009 \\ \cline{2-6}
2000-12-14/15 & MOS1(FF) & 74000 & 0.15-8.0 & 17350(1.6\%) & 0.230$\pm$0.003 \\ \cline{2-6}
81.4  ksec        & MOS2(FF) & 74250 & 0.15-8.0 & 18700(1.6\%) & 0.247$\pm$0.003 \\ \hline \hline
Geminga & pn(SW) & 55000 & 0.15-8.0 & 52850(5.4\%) & 0.909$\pm$0.008 \\ \cline{2-6}
2002-04-05 & MOS1(FF) & 76900 & 0.15-8.0 & 10170(2.2\%) & 0.129$\pm$0.002 \\ \cline{2-6}
103.3  ksec    & MOS2(FF) & 77400 & 0.15-8.0 & 11300(2.4\%) & 0.142$\pm$0.002 \\ 
\hline \hline
\end{tabular}
\end{small}
\tablenotetext{a}{SW: Small Window; Ti: Fast Timing; FF: Full Frame}
\tablecomments{Starting from the left, the column
report (1) the target name, the date of the observation and the total time
span (in ksec) of the observation; (2) the detector and its readout 
mode; (3) the 
good time (in sec) of the observation; 
(4) the energy range considered (in keV); 
(5) the overall
number of counts
in the source extraction region (see Sect.~\ref{srcbkg}) and
the fraction of background events
in the specified energy range; (6) the background-subtracted count rate.}
\end{center}
\end{table}

\clearpage

\begin{table}
\begin{center}
\caption{\label{timingres} Results of Timing analysis.} 
\begin{tabular}{ccc}
\hline \hline
Pulsar & P Observed (ms) & P expected (ms) \\ \hline
PSR B0656+14 & 384.9029(2) &  384.90300043(5) \\ \hline 
PSR B1055-52 & 197.111812(5) & 197.111809432(8) \\ \hline 
Geminga  & 237.1012(1) & 237.1012153(1) \\ \hline \hline
\end{tabular}
\tablecomments{For each target, the best period,
as computed from the EPIC X-ray data, is shown (column 2) together with 
the value expected on the basis of the extrapolation of published ephemeris:
Kern et al.(2003) for PSR B0656+14, the ATNF data pulsar 
archive (http://www.atnf.csiro.au/research/pulsar/psr/archive/)
for PSR B1055-52 and Jackson at al.(2002) for Geminga. The uncertainty quoted
between parentheses refers to the least significant digit.}
\end{center}
\end{table}

\clearpage

\begin{table}
\begin{center}
\caption{\label{spectralres} Results of phase-integrated spectroscopy.} 
\begin{tabular}{l|ccc}
\hline \hline
 & PSR B0656+14 & PSR B1055-52 & Geminga\tablenotemark{a} \\ \hline
N$_H$ (10$^{20}$ cm$^{-2}$) & 4.3$\pm$0.2  & 2.7$\pm$0.2 & 1.07 (fixed) \\
kT$_{CBB}$ (K) & (6.5$\pm$0.1)$\times10^5$ & 
  (7.9$\pm$0.3)$\times10^5$ & (5.0$\pm$0.1)$\times10^5$  \\
R$_{CBB}$ (km)  & 20.9$^{+2.7}_{-3.8}$ & 12.3$^{+1.5}_{-0.7}$ & 8.6$\pm$1.0 \\
kT$_{HBB}$ (K)  & (1.25$\pm$0.03)$\times10^6$ & 
  (1.79$\pm$0.06)$\times10^6$ & (1.9$\pm$0.3)$\times10^6$ \\
R$_{HBB}$ (m) & 1800$\pm$150 & 460$\pm$60 & 40$\pm$10 \\
$\Gamma$ & 2.1$\pm$0.3 & 1.7$\pm$0.1 & 1.7$\pm$0.1 \\
I$_{PL}$ (ph cm$^{-2}$ s$^{-1}$ keV$^{-1}$ @ 1 keV) & 4.3$^{+0.6}_{-1.5}\times10^{-5}$  & 1.9$^{+0.3}_{-0.2} \times10^{-5}$ & 6.7$\pm$0.7$\times 10^{-5}$ \\
F$_{0.2-8 keV}$\tablenotemark{b} (erg cm$^{-2}$ s$^{-1}$) & 1.05$\times 10^{-11} $ & 2.2$\times 10^{-12}$ & 2.3$\times 10^{-12}$ \\
L$_{PL}$\tablenotemark{c} (erg s$^{-1}$) & 1.8$\times 10^{30}$ & 
  8.1$\times 10^{30}$ & 1.2$\times 10^{30}$ \\
L$_{CBB}$\tablenotemark{d} (erg s$^{-1}$) & 5.8$\times 10^{32}$ & 4.4$\times 10^{32}$ & 3.2$\times 10^{31}$ \\
L$_{HBB}$\tablenotemark{e} (erg s$^{-1}$) & 5.7$\times 10^{31}$ & 1.6$\times 10^{31}$ & 1.6$\times 10^{29}$ \\
Norm$_{pn}$\tablenotemark{f} & Ti:1 (fixed); SW:1.05 & 1 (fixed) & - \\
Norm$_{MOS1}$\tablenotemark{f} & 0.96 & 0.98 & - \\
Norm$_{MOS2}$\tablenotemark{f} & - & 1.07 & - \\
$\chi^{2}$/dof & 1.11 & 1.02 & 1.19 \\
dof & 368 & 327 & 73 \\ \hline \hline
\end{tabular}
\tablenotetext{a}{Results from Caraveo et al.(2004)}
\tablenotetext{b}{Observed flux,  0.2-8 keV}
\tablenotetext{c}{0.5-10 keV luminosity of power law component}
\tablenotetext{d}{Bolometric luminosity of the Cool blackbody component}
\tablenotetext{e}{Bolometric luminosity of the Hot blackbody component}
\tablenotetext{f}{Normalization factor to account for calibration differences;
Ti: Fast Timing mode; SW: Small Window mode}
\tablecomments{To compute luminosities we assumed a distance of 288 pc for
PSR B0656+14 (Brisken et al., 2003), of 750 pc for PSR B1055-52 
(Kramer et al., 2003) and of 157 pc for Geminga (Caraveo et al., 1996)}
\end{center}
\end{table}

\clearpage

\clearpage

\clearpage

\clearpage

\clearpage

\clearpage

\clearpage

\clearpage

\clearpage

\renewcommand{\thefigure}{\arabic{figure}\alph{subfigure}}

\setcounter{subfigure}{1}
\begin{figure}
\vspace{-4cm}
\includegraphics[angle=0,width=15cm]{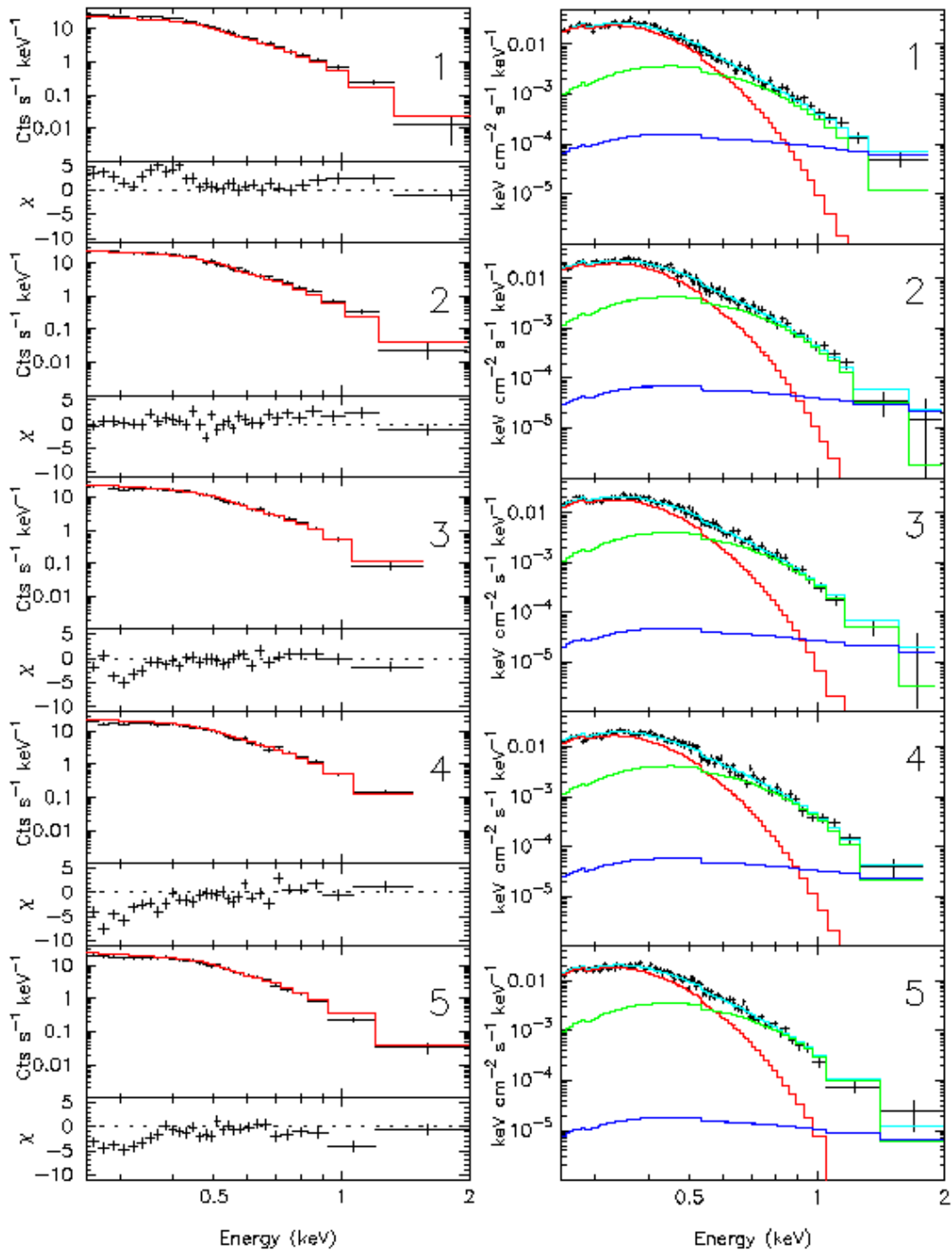}
\vspace{-1cm}
\caption{\label{phaseres_0656_a} Phase-resolved spectra of PSR B0656+14.
Photons have been selected in correspondence of the phase intervals marked
in Fig.~\ref{lc_0656}. Phase intervals from 1 to 5 are shown here;
phase intervals 6-10 are shown in Fig.~\ref{phaseres_0656_b}.
Upper panels on the left column present, for each
phase interval, the observed spectrum
(data points) compared to the 
best fit model (solid line) of the phase-integrated spectrum (upper plots); 
lower panels
show the difference  between data and such model in units of statistical
errors.
To ease the visibility of the deviations 
of phase-resolved
spectra from the averaged spectrum template, spectral channels
have been rebinned. 
Panels on the right column show, for each phase interval, the unfolded
spectrum together with its best fit model. The model components are also
plotted (color code as in Fig.~\ref{spectrum_0656}).
An animated version of Figures 9a/b, 10a/b and 11a/b is available at
http://www.mi.iasf.cnr.it/$\sim$deluca/3musk/
}
\end{figure}

\clearpage

\addtocounter{figure}{-1}
\addtocounter{subfigure}{1}
\begin{figure}
\includegraphics[angle=0,width=15cm]{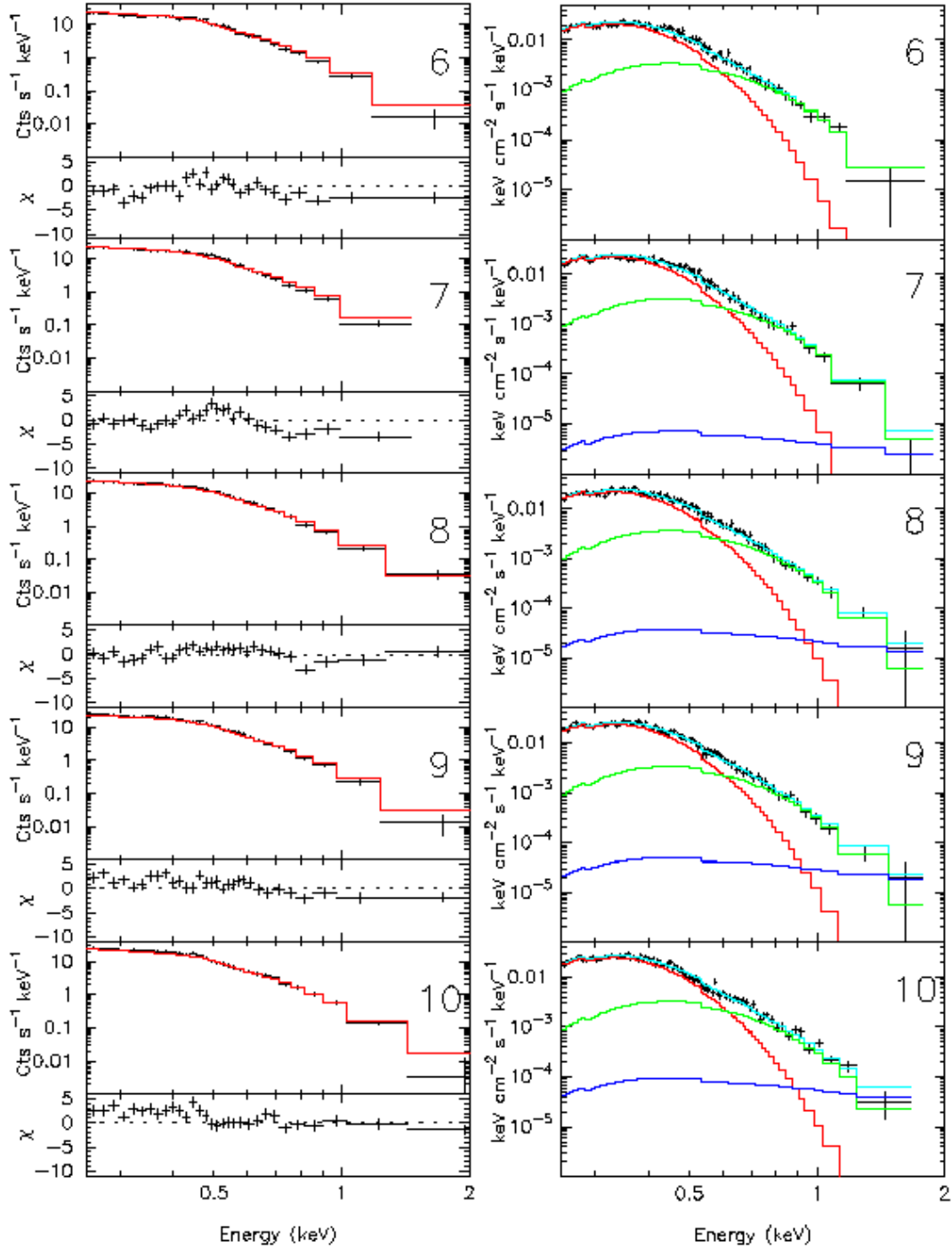}
\caption{\label{phaseres_0656_b} Phase-resolved
spectra of PSR B0656+14. Same as Fig.~\ref{phaseres_0656_a},
phase intervals 6-10 are shown.}
\end{figure}

\clearpage

\addtocounter{subfigure}{-1}
\begin{figure}
\vspace{-2cm}
\includegraphics[angle=0,width=15cm]{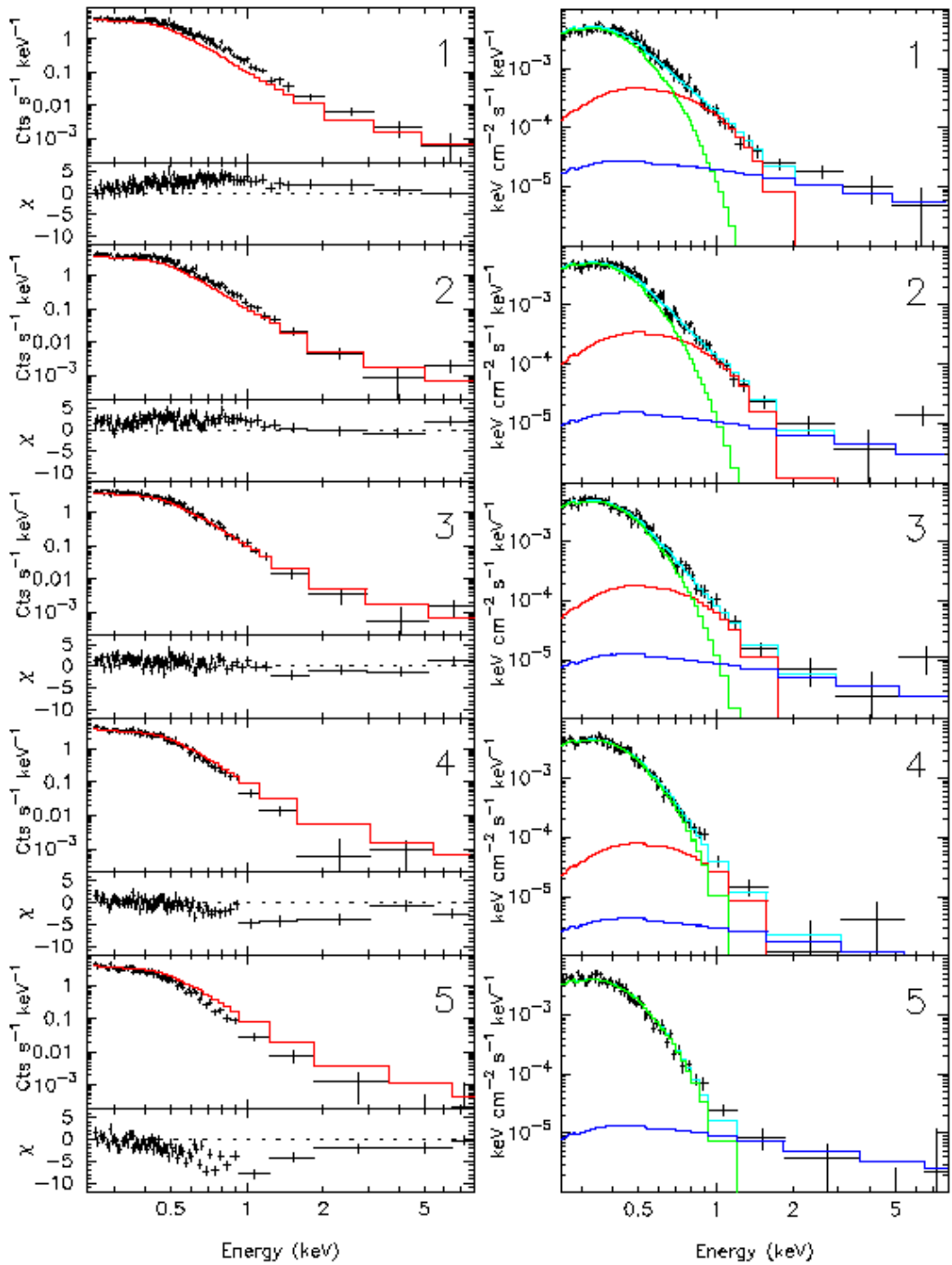}
\caption{\label{phaseres_1055_a}
Phase-resolved spectra of PSR B1055-52.  Phase intervals 
1-5 (according to
the notation of Fig.~\ref{lc_1055}) are displayed.
See caption to 
Fig.~\ref{phaseres_0656_a} for explanations.
Data have not been rebinned. Note the much higher deviations of the
phase-resolved spectra wrt. the average one.}
\end{figure}

\clearpage

\addtocounter{figure}{-1}
\addtocounter{subfigure}{1}

\begin{figure}
\includegraphics[angle=0,width=15cm]{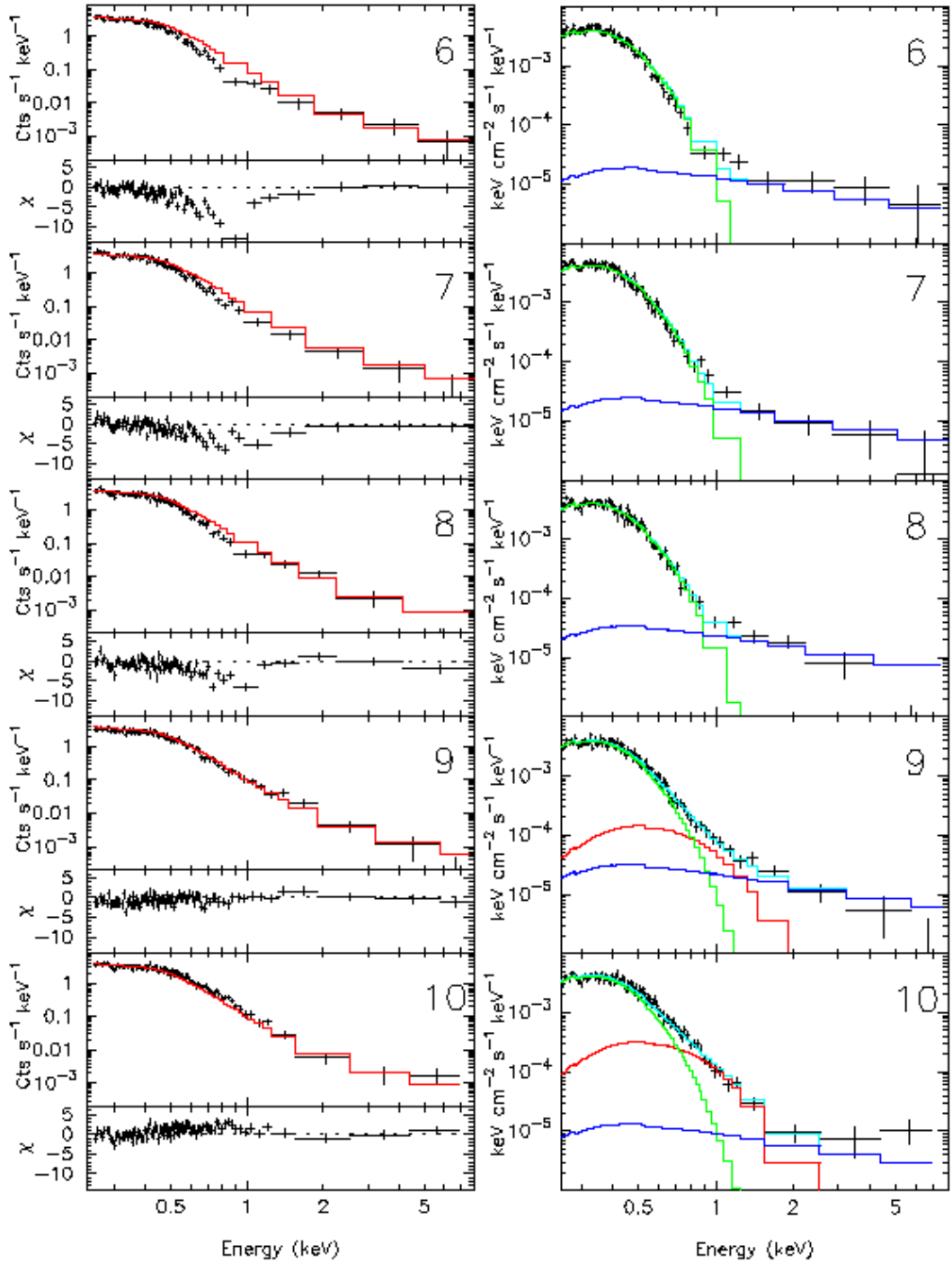}
\caption{\label{phaseres_1055_b} Same as Fig.~\ref{phaseres_1055_a},
phase intervals 6-10 are shown.}
\end{figure}

\clearpage
\addtocounter{subfigure}{-1}

\begin{figure}
\includegraphics[angle=0,width=15cm]{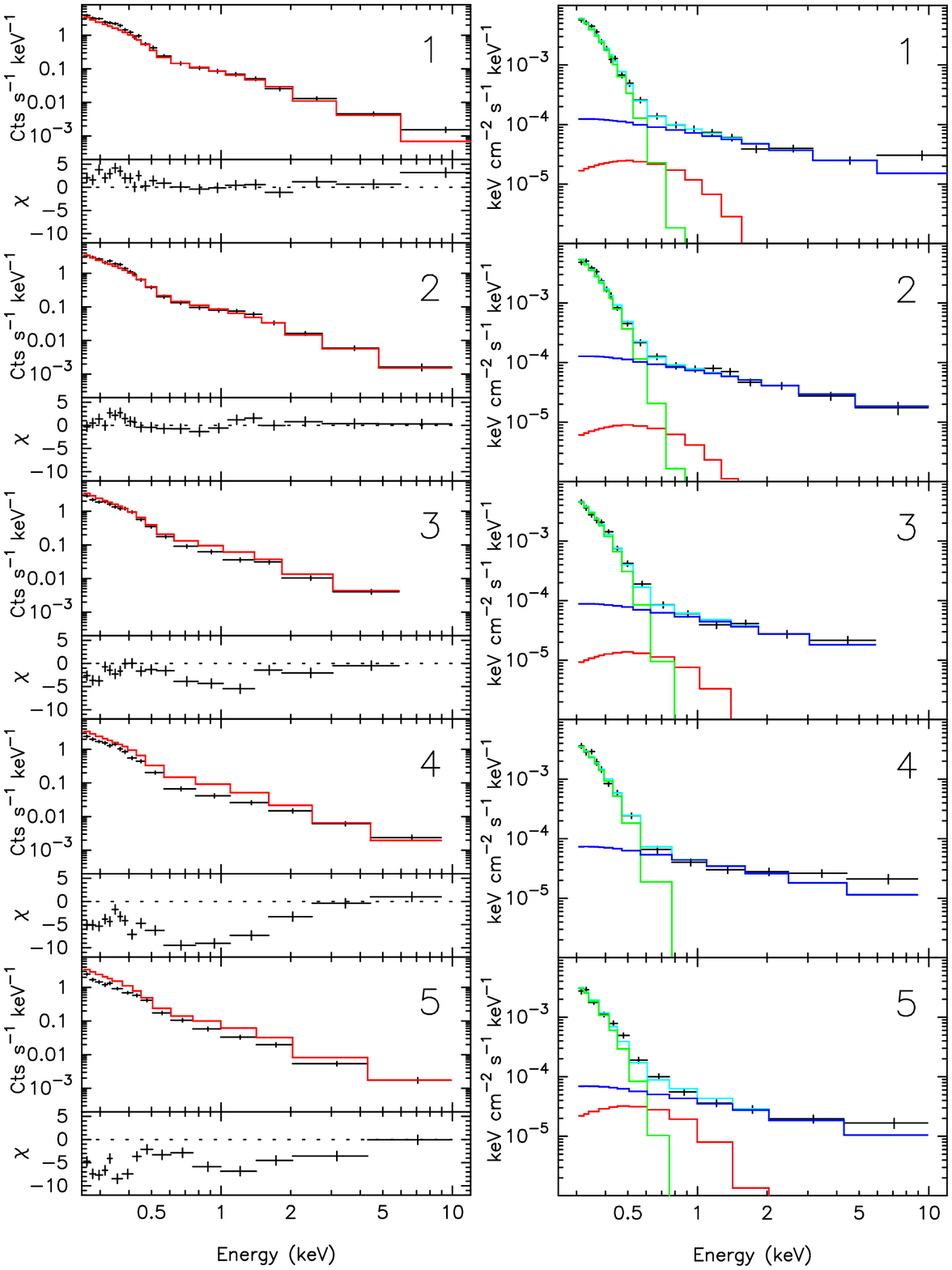}
\caption{\label{phaseres_geminga_a} Phase-resolved spectra of Geminga.
Phase intervals 1-5 (see Fig.~\ref{lc_geminga}) are shown here.
See caption to 
Fig.~\ref{phaseres_0656_a} for explanations.
The figure is adapted from Caraveo et al.(2004), according to 
the phase numbering and color code adopted in this work. }
\end{figure}

\clearpage
\addtocounter{figure}{-1}
\addtocounter{subfigure}{1}

\begin{figure}
\includegraphics[angle=0,width=15cm]{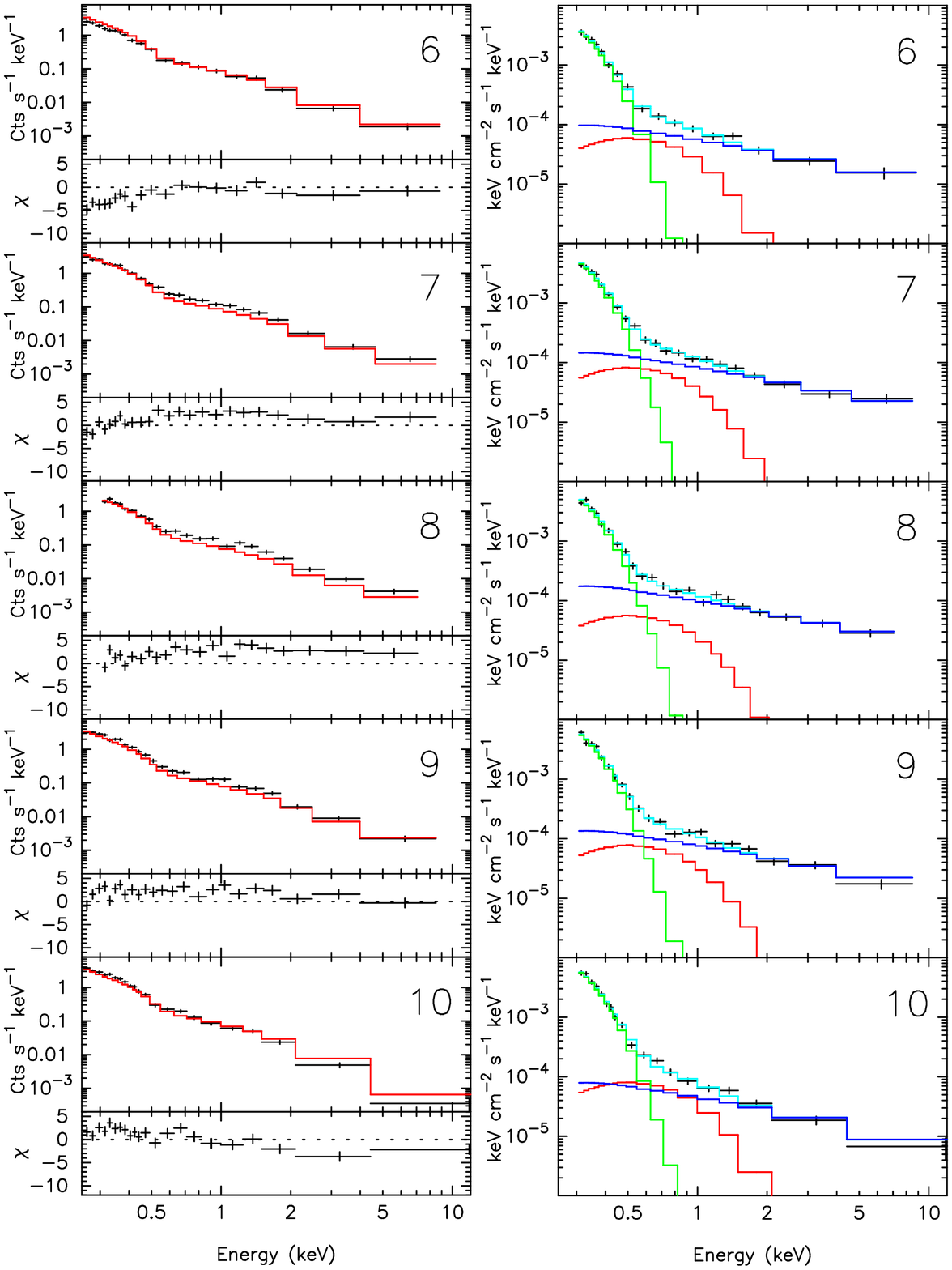}
\caption{\label{phaseres_geminga_b}
Same as Fig.~\ref{phaseres_geminga_a}, phase intervals
6-10 are displayed.}
\end{figure}

\clearpage

\renewcommand{\thefigure}{\arabic{figure}}

\begin{figure}
\includegraphics[angle=-90,width=15cm]{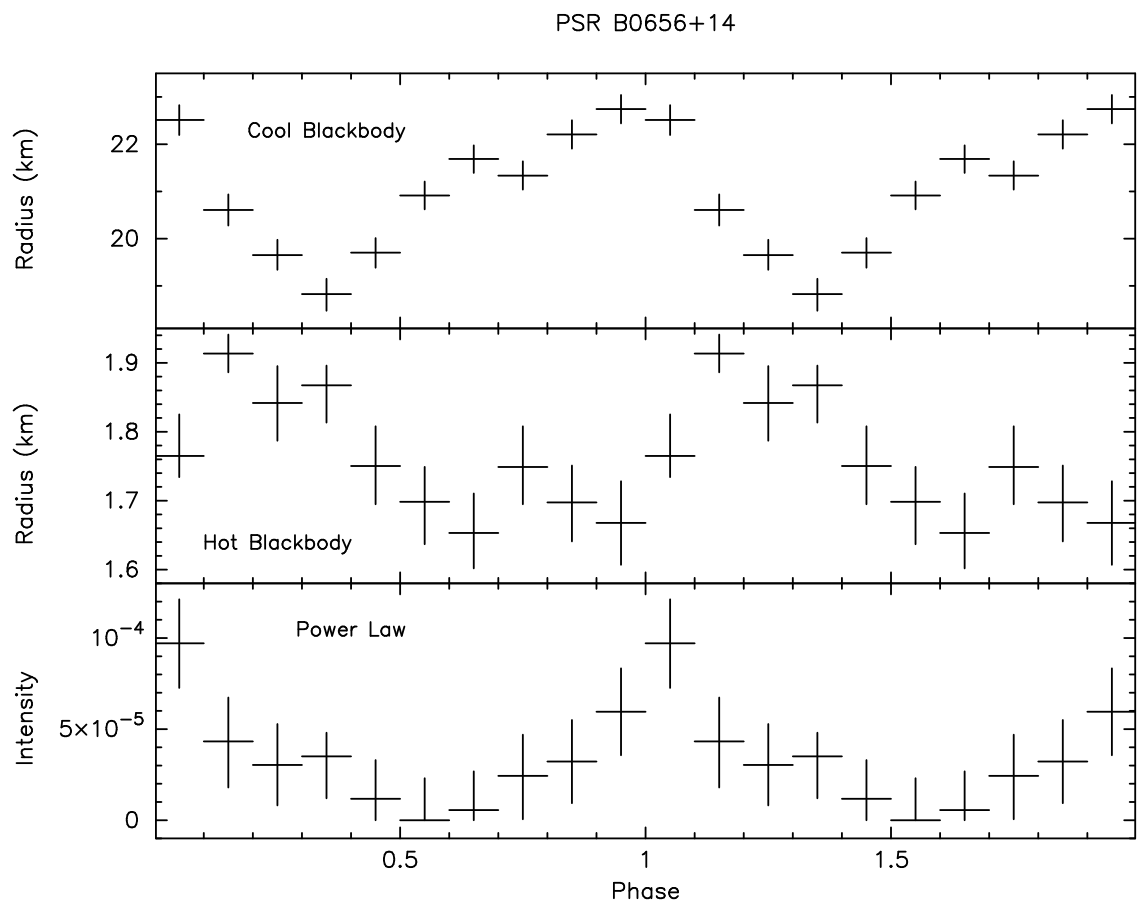}
\caption{\label{parms_0656} The parameters 
best fitting the phase-resolved spectra of PSR B0656+14
are plotted as a function of the pulsar
phase defined as in Fig.~\ref{lctot}.
Both cool and hot blackbody 
emitting surfaces evolve throughout the pulsar phase following
a sinusoidal pattern showing an overall $\sim$10\%
modulation (wrt. the average values) on the emitting radii value. 
Note the marked anti-correlation
between panel 1 and panel 2 with the hot blackbody peaking
in correspondence of the cool blackbody minimum.
The power law component has a different phase trend
wrt. the thermal components, with a single,
sharper peak.
}
\end{figure}

\clearpage

\begin{figure}
\includegraphics[angle=-90,width=15cm]{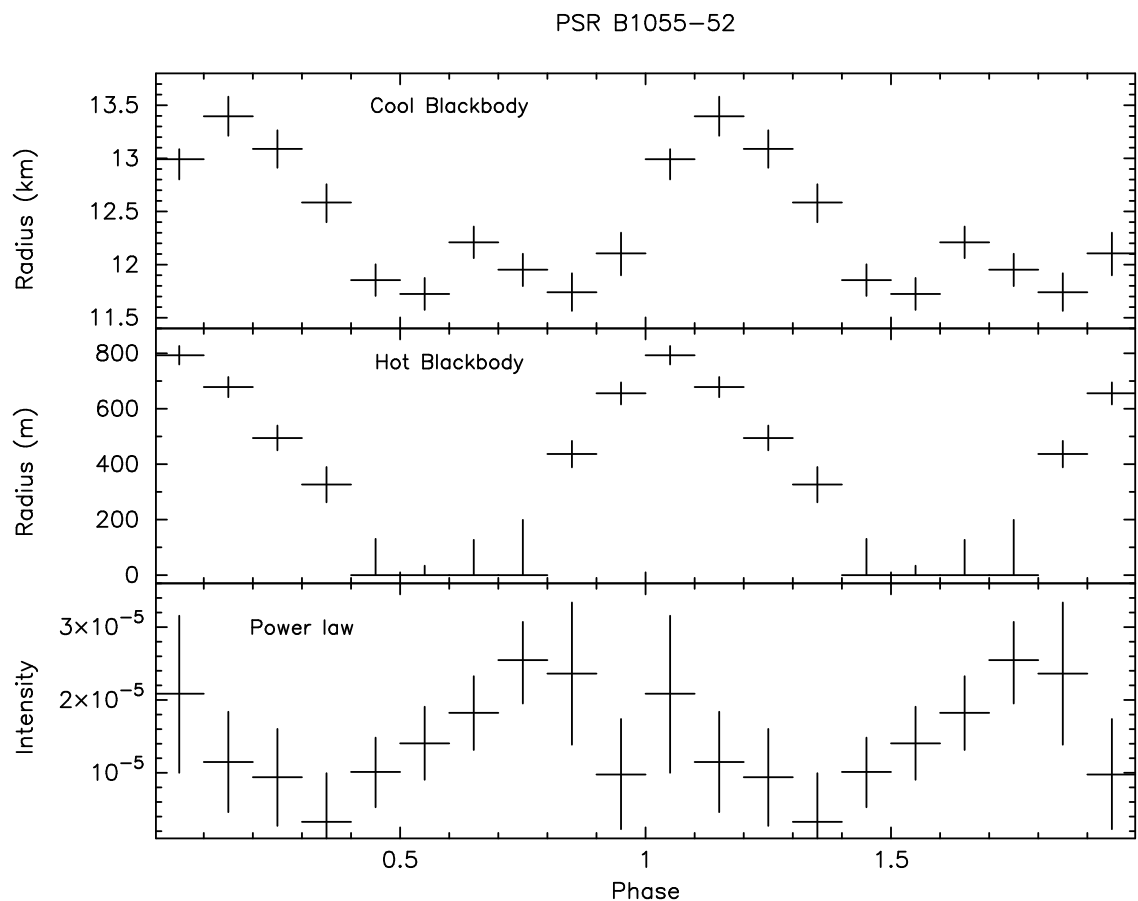}
\caption{\label{parms_1055} Same as Fig.\ref{parms_0656}, for
PSR B1055-52. While the cool blackbody emitting radius
shows a $<$10\%
modulation (wrt. the average value), the hot blackbody component
show a dramatic, 100\% modulation 
since its contribution is null in 4 of the 10 phase intervals.
Note that for PSR B1055-52 the two thermal
components have a similar time evolution, with a phase shift as low as 
$\sim$0.1. The power law component has a different modulation; it is difficult
to assess wether its profile is single-peaked or double-peaked, owing
to the lower signal to noise in the high energy portion of the spectra.}
\end{figure}

\clearpage

\begin{figure}
\includegraphics[angle=-90,width=15cm]{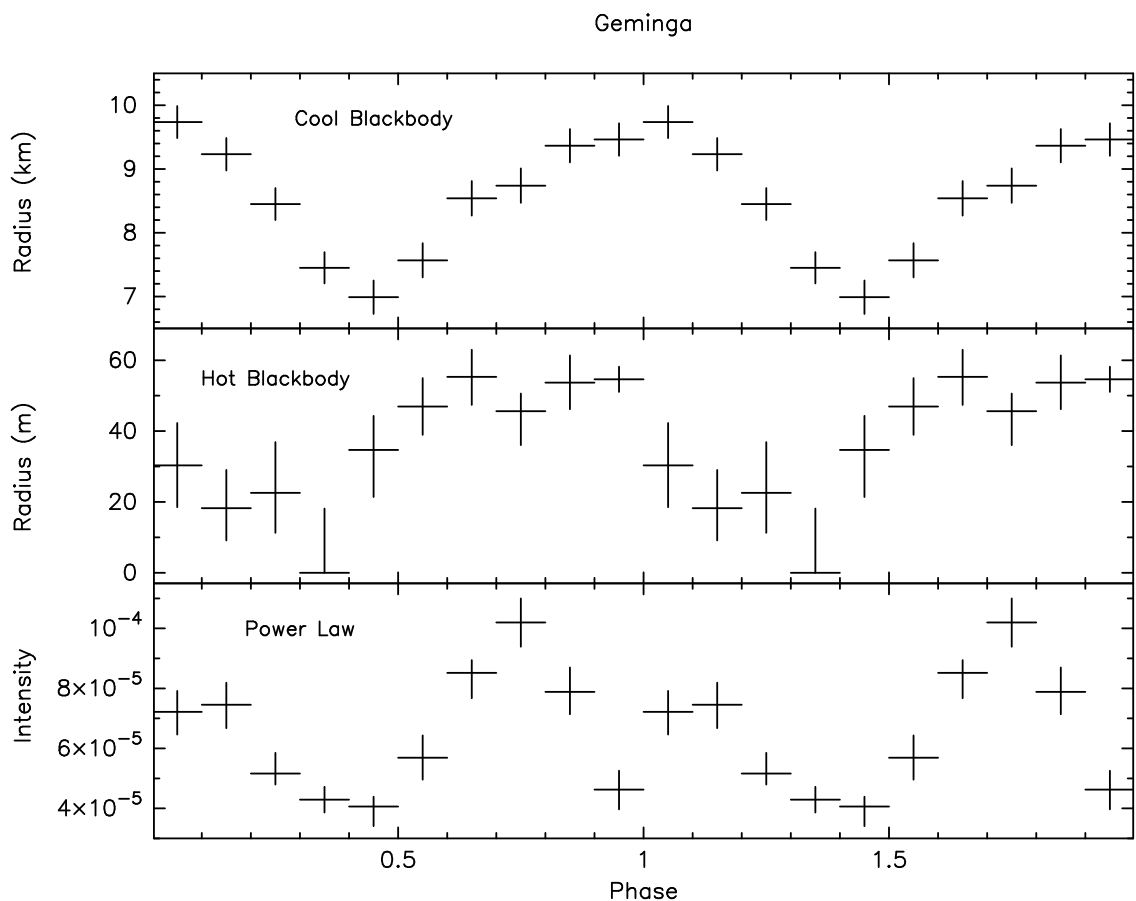}
\caption{\label{parms_geminga} Same as Fig.\ref{parms_0656} and 
Fig.\ref{parms_1055} for the case of Geminga. The figure is adapted from
Caraveo et al.(2004) (see their
Figure 4), according to the choice of phase adopted in this work.
The cool blackbody component shows a $\sim15$\% 
modulation (wrt. the average value) of its emitting
radius, with a sinusoidal profile. Conversely, the hot blackbody is
100\% modulated, and disappears for 1/10 of the pulsar period.
The power law component has a remarkably different, double-peaked phase
profile and shows a significant unpulsed fraction.
}
\end{figure}


\begin{references}
\reference{} Arons, J. \& Scharlemann, E.T., 1979, ApJ 231, 854
\reference{} Becker, W., Weisskopf, M.C., Tennant, A.F., et al., 2004, ApJ
615, 908 
\reference{} Becker, W. \& Aschenbach, B.M., 2002, Proceedings of the 270. 
WE-Heraeus Seminar on Neutron Stars, Pulsars, and Supernova Remnants. 
Edited by W. Becker, H. Lesch, and J. Trümper. Garching bei München:
Max-Plank-Institut für extraterrestrische Physik, p.64
\reference{} Becker, W. \& Tr\"umper, J., 1997, A\&A 326, 682
\reference{} Bertsch et al., 1992, Nature 357, 306
\reference{} Bignami, G.F., Caraveo, P.A., Paul, J.A., Salotti, L., Vigroux,
L., 1987, ApJ 319, 358
\reference{} Bignami, G.F., Caraveo, P.A., Lamb, R.C., 1983, ApJ 272, L9
\reference{} Brisken, W.F.,  Thorsett, S.E., Golden, A., Goss, W.M., 2003
ApJ 593, L89
\reference{} Burwitz, V. et al., 2004, SPIE 5165, 123
\reference{} Caraveo, P.A., et al., 2003, Science 301, 1345
\reference{} Caraveo, P.A., De Luca, A., Mereghetti, S., Pellizzoni, A.,
Bignami, G.F., 2004, Science, 305, 376
\reference{} Caraveo, P.A., Bignami, G.F., Mignani, R.P., Taff, L., 1996,
ApJ 461, L91
\reference{} Caraveo, P.A., Bignami, G.F., Mereghetti, S., 1994 ApJ 422, L87
\reference{} Cordova, F.A., Middleditch, J., Hjellming, R.M., Mason, K.O., 
1989, ApJ 345, 451
\reference{} Cheng, A.F. \& Helfand, D.J., 1983, ApJ 271, 271
\reference{} De Luca, A., \& Molendi, S., 2004, A\&A 419, 837
\reference{} Everett, J.E. \& Weisberg, J.M., 2001, ApJ 553, 341
\reference{} Fierro, J.M., et al., 1993, ApJ 413, L27 
\reference{} Finley, J.P., Oegelman, H., Kiziloglu, U., 1992, ApJ 394, L21
\reference{} Greenstein, G. \& Hartke, G.J., 1983, ApJ 271, 283
\reference{} Greiveldinger, C., et al., 1996, ApJ 465, L35 
\reference{} Halpern, J.P. \& Holt, S.S., 1992, Nature 357, 222
\reference{} Halpern, J.P. \& Ruderman, M., 1993, ApJ 415, 286
\reference{} Halpern, J.P. \& Wang, F.Y.-H., 1997, ApJ 477, 905
\reference{} Harding, A.K. \& Muslimov, A.G., 2002, ApJ 568, 862
\reference{} Harding, A.K. \& Muslimov, A.G., 1998, ApJ 500, 862
\reference{} Jackson, M.S., Halpern, J.P., Gotthelf, E.V., 2002, ApJ 538, 935
\reference{} Kern, B., Martin, C., Mazin, B., Halpern, J.P., 2003
ApJ 597, 1049
\reference{} Kirsch, M., on behalf of the EPIC calibration team, 2004, EPIC
Calibration Status, Document CAL-TN-0018-2-3, available from http://xmm.vilspa.esa.es/external/xmm\_sw\_cal/calib/index.shtml
\reference{} Kramer, M., et al., 2003, MNRAS 342, 1299
\reference{} Lattimer, J.M. \& Prakash, M., 2001, ApJ 550, 426
\reference{} Lyne, A.G. \& Manchester, R.N., 1988, MNRAS 234, 477
\reference{} Marshall, H.L. \& Schulz, N.S., 2002, ApJ 574, 377
\reference{} Mignani, R.P., De Luca, A., Caraveo, P.A., 2004, in ``Young 
Neutron Stars and Their Environments'', Proc. IAU Symp.218, Editors Camilo, F.,
Gaensler, B.M., ASP, p.391
\reference{} Mignani, R.P., Caraveo, P.A., Bignami, G.F., 1997, ApJ 474, L51
\reference{} Oegelman, H., 1995, The Lives of the Neutron Stars. 
Proceedings of the NATO Advanced Study Institute on the Lives of the 
Neutron Stars, Editors M.A. Alpar, U. Kiziloglu, J. van Paradijs;
Publisher, Kluwer Academic, Dordrecht, The Netherlands, Boston, Massachusetts
\reference{} Oegelman, H. \& Finley, J.P.,  1993, ApJ 413, L31
\reference{} Page, D., 1995, ApJ 442, 273
\reference{} Pavlov, G.G., Zavlin, V.E., Sanwal, D., 2002, 
Proceedings of the 270. 
WE-Heraeus Seminar on Neutron Stars, Pulsars, and Supernova Remnants. 
Edited by W. Becker, H. Lesch, and J. Trümper. Garching bei München:
Max-Plank-Institut für extraterrestrische Physik, p.273
\reference{} Possenti, A., Mereghetti, S., Colpi, M., 1996, A\&A 313, 565
\reference{} Psaltis, D., Oezel, F., DeDeo, S., 2000, ApJ 544, 390
\reference{} Ramanamurthy, P.V., Fichtel, C.E., Kniffen, D.A., Sreekumar, P., 
Thompson, D.J., 1996, ApJ 458, 755
\reference{} Rankin, J.M., 1993, ApJS 85, 145
\reference{} Ruderman, M. \& Sutherland, P.G., 1975, ApJ 196, 51
\reference{} Ruderman, M., 2003, 4th AGILE Science Workshop, 
``X-ray and Gamma-ray Astrophysics of Galactic Sources'', astro-ph/0310777
\reference{} Str\"uder, L., et al., 2001, A\&A 365, L18
\reference{} Thompson, D.J., et al., 1999, ApJ 516, 297 
\reference{} Turner, M.J.L., et al., 2001, A\&A 365, L27
\reference{} Zavlin, V.E. \& Pavlov, G.G., 2002, Proceedings of the 270. 
WE-Heraeus Seminar on Neutron Stars, Pulsars, and Supernova Remnants. 
Edited by W. Becker, H. Lesch, and J. Trümper. Garching bei München:
Max-Plank-Institut für extraterrestrische Physik, p.263
\reference{} Zavlin, V.E., Pavlov, G.G., Shibanov, Y.A., 1996, A\&A 315, 141
\reference{} Zavlin, V.E., Shibanov, Y.A., Pavlov, G.G., 1995, Astronomy 
Letters 21, 149

\end{references}
\end{document}